\documentclass[11pt]{article}

\usepackage[latin1]{inputenc}
\usepackage{amsmath}
\usepackage{amsfonts}
\usepackage{amssymb}
\usepackage{latexsym}
\usepackage{mathrsfs}
\usepackage{graphicx}
\usepackage{pstricks}
\usepackage{color}
\usepackage[colorlinks=true, citecolor=blue]{hyperref}

\newcommand{\C}{\mathbb{C}}

\newcommand{\be}{\begin{equation}\label}
\newcommand{\ee}{\end{equation}}
\newcommand{\bea}{\begin{eqnarray}\label}
\newcommand{\eea}{\end{eqnarray}}

\newtheorem{remark}{Remark}[section]

\numberwithin{equation}{section}

\begin{document}

\title{Light cones in relativity: Real, complex and virtual, with applications}

\author{T.M. Adamo \\ \textit{The Mathematical Institute} \\ \textit{University of Oxford} \\ \textit{24-29 St Giles'} \\  \textit{Oxford, OX1 3LB, U.K.} \\ \texttt{adamo@maths.ox.ac.uk}
\and E.T. Newman \\ \textit{Department of Physics \& Astronomy} \\ \textit{University of Pittsburgh} \\ \textit{3941 O'Hara St.} \\ \textit{Pittsburgh, PA 15260, U.S.A.} \\ \texttt{newman@pitt.edu}}

\maketitle

\abstract{We study geometric structures associated with shear-free null geodesic
congruences in Minkowski space-time and asymptotically shear-free null
geodesic congruences in asymptotically flat space-times. We show how in both the flat and
asymptotically flat settings, complexified future null infinity, $\mathfrak{I}_{\C}^{+}$, acts as a \textquotedblleft holographic screen,\textquotedblright\ interpolating
between two dual descriptions of the null geodesic congruence. One
description constructs a complex null geodesic congruence in a complex space-time whose
source is a complex world-line; a virtual source as viewed from the
holographic screen. This complex null geodesic congruence intersects the real asymptotic boundary
when its source lies on a particular open-string type structure in the complex space-time. The other description constructs a real, twisting, shear-free or asymptotically shear-free null geodesic congruence in the real space-time, whose source (at least in Minkowski space) is in general a closed-string structure: the caustic set of the congruence. Finally we show that virtually all of the interior space-time physical quantities that are identified at null infinty $
\mathfrak{I}^{+}$, (center of mass, spin, angular momentum, linear momentum,
force) are given kinematic meaning and dynamical descriptions in terms of
the complex world-line.}

\pagebreak

\tableofcontents

\section{Introduction}

\bigskip In this paper, we describe some interesting structures based in
classical special and general relativity which bear some resemblance to dualities known
as \textquotedblleft holographic\textquotedblright\ dualities which have
emerged elsewhere in theoretical physics over the past decades (c.f., \cite{'tHooft:1974, 'tHooft:1993gx, Bousso:2002}). \ Though
these holographic dualities usually involve the use of highly non-classical
machinery such as supersymmetry or string theory, most famously in the case of the
AdS/CFT correspondence \cite{Maldacena:1997re, Witten:1998qj, Aharony:1999ti}, we emphasize that our discussion here will use no
such tools; we work entirely in the context of classical four-dimensional
Lorentzian space-time. \ The structures we are interested in emerge
naturally from the study of light cone foliations (and their generalization
to asymptotically shear-free null geodesic congruences) in space-time, and
have simply been overlooked in prior research. \ Although it would certainly
be presumptuous for us to suggest that our work here has any true connection
with holographic duality as it is known to most theoretical physicists, we
do find a holographic screen, open and closed classical strings and other
suggestive objects; all of which can be given real physical meaning in
four-dimensional space-time.\ 

Specifically, it is the purpose of this note to first explore the properties
of ordinary \textquotedblleft run of the mill\textquotedblright\ light-cones
and then turn to their generalization via complex and virtual light cones in
four-dimensional Lorentzian space-times. More precisely, we study the
properties of light-cones and their complex generalizations in both
Minkowski space and in asymptotically flat (vacuum and Einstein-Maxwell) 
space-times in the neighborhood of future null infinity. \ This is followed by a discussion of physical applications of these ideas and constructs. \ These complex light
cones are first applied to the structure of \textit{real} Maxwell fields in 
\textit{real} Minkowski space. The complex cones in flat space-time are then generalized and applied to study equations of motion in general relativity. Along the way we point out a pretty duality between the complex light-cones and real shear-free, but
twisting, null geodesic congruences.

The first issue raised comes from the simple question: In Minkowski space,
avoiding or ignoring their apex, what are the geometric properties that a
set of null geodesics must have in order to be a light cone? How can they be
determined to be light cones even far from their apex? The answer is simple:
First of all, the family of relevant null geodesics (the light-cone
generators) must be (null) surface forming; they must lie on a null surface
and thus have vanishing \textquotedblleft twist.\textquotedblright\ As a
result, these surfaces must be foliated by null geodesics whose tangent
vectors are determined by the gradient of the surface. Second, they must
have vanishing shear and non-vanishing divergence. (The plane null surfaces
can be thought of as light-cones but with their apex at infinity; we ignore
this case.) From this requirement, light cones posses the topology of $%
S^{2}\times \mathbb{R}.\ $\ For us, the most relevant feature is their 
\textit{vanishing shear}. Even far from their apex (i.e., even at future
null infinity, $\mathfrak{I}^{+}$)$\ $on the $S^{2}\ $portion of the
surface, if the shear vanishes then it has an apex and the surface is a
light-cone. \ This case can be generalized from an individual light-cone to
a family of light-cones: if Minkowski space is (partially) foliated by a
null geodesic congruence (NGC), can we tell at $\mathfrak{I}^{+}\ $that the
geodesics all focus to a time-like world-line in the interior? \ The answer
again is simple: If the congruence has vanishing twist and shear at $%
\mathfrak{I}^{+}\ $and non-vanishing divergence, and furthermore has no
members lying tangent to $\mathfrak{I}^{+}\ $itself (the regularity
condition), then there is a one-parameter family of light-cones and a
time-like world-line to which the NGC converges.

The main goal of this work is to investigate and analyze how this asymptotic
description of light cones can be generalized, and what applications to physics it might have. The generalization will be in two distinctly different but related directions.

First of all, in the context of Minkowski space, we define and describe 
\textit{complex light cones}. They will be determined solely from the
properties of specific sets of null directions at complexified null infinity
($\mathfrak{I}_{\mathbb{C}}^{+}$), the analytic continuation of Penrose's future null infinity, $\mathfrak{I}^{+}$. \ These complex null directions, normal
to specific slices of $\mathfrak{I}_{\mathbb{C}}^{+},\ $define - by
following them backwards in time - complex null geodesics (and complex
light-cones) which converge to points in complex Minkowski space \cite{Adamo:2009vu, Adamo:2009ru}. In general
there will be a subset of points in complex Minkowski space where one (or
more) of its light-cone generators intersects real Minkowski space at real $%
\mathfrak{I}^{+}$. If instead of the complex light-cone of a single point in
complex Minkowski space we take a complex analytic \textquotedblleft
time-like\textquotedblright\ (to be defined) world-line parametrized by the
complex parameter $\tau $,$\ $we would have a two-real-dimensional set of
complex points (from the real and imaginary parts of $\tau $)$\ $and their
light cones. We show that for any fixed value of the real part there is a
one-dimensional set of points such that the envelope formed by their individual light
cones intersects real $\mathfrak{I}^{+}\ $on an $S^{2}\ $slice. As the real
parameter changes we obtain a one parameter family of real slicings of $%
\mathfrak{I}^{+}$.

At this point, an interesting duality emerges. On one hand, if we start from
each of these slices and move backwards into the complex space along the
complex null directions, these trajectories converge to an imaginary line
segment in the complex space. On the other hand, there is a dual method
(described later) for following null geodesics from the slices back into the
real space-time; this yields a \textit{real} shear-free, but twisting, null
geodesic congruence. It is precisely this twist which links the two
pictures: the \textquotedblleft distance\textquotedblright\ of the complex
world-line from the real Minkowski space-time in the first picture is a
measure of the twist of the real congruence in the latter picture. The
caustic set of the real (dual) congruence is (in general) a closed curve
moving in real time; something analogous to a classical closed string \cite{Adamo:2009ru}.

The extension of these ideas to asymptotically flat Einstein space-times
initially seems to be impossible. Standard light cones from any given
space-time point will undergo such distortions from the curvature of the
space-time itself that little or no memory of their origin will remain when they
arrive at $\mathfrak{I}^{+}$. Nevertheless, we can consider the
possibility of using the procedure that was succesful in the Minkowski
space-time case$\ $by asking for null geodesic congruences in the
neighborhood of $\mathfrak{I}^{+}\ $that are shear-free and non-twisting. In
the general asymptoically flat case, \textit{shear-free }null geodesic
congruences do not exist - but there are always null geodesic congruences
that are \textit{asymptotically shear-free} in the neighborhood of $%
\mathfrak{I}^{+}$. \ Unfortunately, to use this idea effectively again
entails the analytic extension of the space-time a small distance into the
complex. \ Working on the complexification of $\mathfrak{I}^{+}\ $(i.e., on $%
\mathfrak{I}_{\mathbb{C}}^{+}$),$\ $there is a construction of complex
slices or \textquotedblleft cuts\textquotedblright\ whose complex null
normals can be used to determine \textit{asymptotically shear-free and
twist-free complex null geodesics} \cite{NT-80, Kozameh:2006zw, Adamo:2009vu}. In fact one can construct a four-complex
dimensional family of such complex cuts which define a four-complex
dimensional manifold frequently referred to as $\mathcal{H}$-space \cite{Newman76, HNPT78, Hspace77}.\footnote{Though it is not used here, we mention that $\mathcal{H}$-space is endowed with a complex Ricci-flat metric with self-dual Weyl
tensor.}  The immediately relevant feature for us is that these complex null
geodesics from each complex \textquotedblleft
cut\textquotedblright\ converge or focus to a point in $\mathcal{H}$-space \cite{Ludvigsen:1980}.
It will be shown later that real structures associated with $\mathcal{H}$%
-space can be found, and that real physics can be interpreted as taking
place in $\mathcal{H}$-space \cite{Kozameh:2008gw, Adamo:2009vu}. The $\mathcal{H}$-space can thus be viewed as
the virtual image space seen\ by looking-backwards\ along complex null
directions from a sphere of points on $\mathfrak{I}_{\mathbb{C}}^{+}$. $\ $%
It is this property that could allow us to refer to $\mathfrak{I}_{\mathbb{C}%
}^{+}\ $as a holographic screen.

The prior discussion of complex Minkowski space (which is a special case of $%
\mathcal{H}$-space) can be extended to $\mathcal{H}$-space. There is a
subset of points in $\mathcal{H}$-space where one (or more) of the light
cone generators (null geodesics) coming from a complex point intersects real
asymptotically flat space-time at real\textit{\ }$\mathfrak{I}^{+}$. If
instead of the complex light cone of a single point in complex $\mathcal{H}$%
-space, we take a complex \textquotedblleft time-like\textquotedblright\ (to
be defined) world-line parametrized by the complex parameter $\tau
=s+i\lambda $,$\ $we would have a two-real-dimensional set of complex points
(from the real and imaginary parts of $\tau $)$\ $and their associated
complex light cones$.\ $For any fixed value of the real part, $s$, there is
a one-dimensional set of points (a finite interval parametrized by $\lambda $%
) such that the envelope formed by their individual light-cones intersects real $%
\mathfrak{I}^{+}\ $on a $S^{2}\ $slice. As the real parameter $s$ changes we
obtain in the $\mathcal{H}$-space a ribbon (the finite interval moving in
\textquotedblleft $s$-time\textquotedblright )\ which could be called a
classical open string; from the null cones of points on this ribbon, we get
a one parameter family of real slicings on $\mathfrak{I}^{+}$. \ All the
information about the ribbon is encoded (holographically) in the one
parameter family of real $\mathfrak{I}^{+}\ $slicings and a null direction
field on $\mathfrak{I}^{+}$. In other words, there is a duality between the
coded information on $\mathfrak{I}^{+}\ $and $\mathcal{H}$-space
information. \ \ A further related duality is that a given complex analytic
world-line in $\mathcal{H}$-space (via its associated ribbon) yields in the
physical space-time (via a complex conjugate action) a \textit{real }%
twisting but asymptotically shear-free null geodesic congruence in the real
space-time.

The question of where this beautiful mathematical structure makes
contact with physical issues does have a simple answer - the details,
however, are rather complicated.

The simple answer is that an (analytic) asymptotically flat Maxwell field in
Minkowski space with non-vanishing total charge generates, in the complex
Minkowski space, a unique complex analytic world-line: the \textit{complex
center of charge} world-line, where the real part describes the standard
center of charge while the ribbon thickness encodes magnetic dipole
information \cite{Newman:2004, Kozameh:2007sk}. \ For the case of asymptotically flat space-times there are two
situations: the vacuum asymptotically flat and the Einstein-Maxwell
asymptotically flat space-times. For the vacuum case there is a unique
complex $\mathcal{H}$-space world-line that contains: from the real part,
the equations of motion for the physical center of mass and, from the ribbon
thickness, the spin angular momentum, with both interpretations arising from
the \textquotedblleft view\textquotedblright\ at infinity, $\mathfrak{I}^{+}$%
\cite{Kozameh:2008gw, Adamo:2009vu}.\ These are (loosely) analogues of measuring the total charge at infinity
via Gauss's law or observing the Bondi energy-momentum vector at infinity.

In Section \ref{2}, the preliminaries, we introduce our notation and results from
earlier investigations that will be needed here. Specifically we first
discuss conventions and notation followed by a description of flat-space
null geodesic congruences. The section ends with a brief summary of
properties of asymptotically flat spaces and their null geodesic
congruences. Section \ref{3} deals with real space-time structures that are
associated with the complex world-lines; first in complex Minkowski space
and then in $\mathcal{H}$-space. In Section \ref{4} we apply the ideas associated
with the complex world-lines to real physical ideas. In particular we show
that a real asymptotically flat Maxwell field (with non-vanishing charge)
determines a complex world-line (the complex center of charge) that carries
information about both the electric and magnetic dipole moments. This
construction is then generalized to asymptotically flat space-times, where
the complex mass dipole moment (the real mass dipole moment plus
\textquotedblleft $i$\textquotedblright\ times the angular momentum)
determines an $\mathcal{H}$-space world-line. The Bianchi identities then
yield kinematic definitions, equations of motion, angular momentum and
conservation laws. All take place in $\mathcal{H}$-space which we interpret
as a virtual image space. This information is coded into the real space-time
by functions on real $\mathfrak{I}^{+}.\ $The results are very reminicent of
ordinary Newtonian dynamical laws of motion.  Though partially a summary of results presented elsewhere in the literature (e.g., \cite{Adamo:2009vu, Kozameh:2008gw, Kozameh:2007sk}), our presentation includes several simplifications and alterations.
In Section \ref{5} we summarize
the earlier discussion and speculate on what meaning and possible future use
there might be to the observations made here.  Appendix \ref{A} provides some background on tensorial spin-$s$ spherical harmonics, which are used throughout this work.

We again stress that the strange results described here lie wholey in
standard four-dimensional classical physics. There is no need - other than
assumed analyticity - to rely on drastic modifications of space-time
properties such as supersymmetry or higher dimensions. The results are here
to be seen and perhaps understood. \ It would have been a cruel god to have
layed down such a pretty scheme and not have it mean something deep.

\section{Foundations: $\mathfrak{I}^{+},\ $Null Geodesic Congruences and
Asymptotic Flatness}
\label{2}

In this section we summarize several of the basic ideas and tools which are
needed in our later discussions.\ The explanations are rather concise and
extensive proofs are omitted.\ In large part, much of what is covered in
this section should be familiar to many or even most workers in general
relativity.

\subsection{Conventions and Notation}

The arena for most of our discussion is the neighborhood of the ``far
(infinite) null future'' of our space-time, (intuitively the end-points of
future directed null geodesics) for both Minkowski space and asymptotically
flat space-times. This region, first defined and studied by Roger Penrose
and referred to as future null-infinity, $\mathfrak{I}^{+}$,$\ $is
constructed by the rescaling of the space-time metric by a conformal factor
which appoaches zero asymptotically, the zero value defining $\mathfrak{I}%
^{+}$ \cite{Penrose:1963, Penrose:1965, Frauendiener:2004}. \ This process leads to the (future-null) boundary being a null
hypersurface for the conformally rescaled metric with topology $S^{2}\times 
\mathbb{R}.$ An easy visualization of the boundary $\mathfrak{I}^{+}$ is as
the past light cone of the point $\boldsymbol{I}^{+}$, future time-like
infinity. A natural coordinatization of $\mathfrak{I}^{+}\ $and its
neighborhood is via a Bondi coordinate system: $(u,r,\zeta ,\bar{\zeta})$. \
\ In this system, $u$, the Bondi time, labels the null surfaces of the
space-time that intersect $\mathfrak{I}^{+}$; $r$ is the affine parameter
along the null geodesics of the constant $u$ surfaces; and $\zeta =e^{i\phi
}\cot (\theta /2)$ is the complex stereographic angle that labels the null
geodesics of $\mathfrak{I}^{+}$, the $S^{2}\ $portion of $\mathfrak{I}^{+}$ \cite{NT-80}.
\ 

In Minkowski space, the Bondi coordinates $(u,\zeta ,\overline{\zeta })$ of $%
\mathfrak{I}^{+}\ $can be constructed from the intersection of the future
null cones of the time-like world-line at the Minkowski space spatial
origin, i.e., from the line, $x^{a}=(t,0,0,0).\ $The cone has the form

\begin{eqnarray*}
x^{a} &=&u_{ret}\delta _{0}^{a}+r\hat{l}^{a}(\zeta ,\overline{\zeta }),\ \ 
\\
\zeta  &=&e^{i\phi }\cot (\theta /2),\ \ u_{ret}=t-r=\sqrt{2}u \\
\hat{l}^{a} &=&\frac{\sqrt{2}}{2(1+\zeta \bar{\zeta})}\left( 1+\zeta \bar{%
\zeta},\ \zeta +\bar{\zeta},-i(\zeta -i\bar{\zeta}),-1+\zeta \bar{\zeta}%
\right)  \\
&=&\frac{\sqrt{2}}{2}(1,\sin \theta \cos \phi ,\sin \theta \sin \phi ,\cos
\theta )
\end{eqnarray*}%
with $(\zeta ,\overline{\zeta })\ $labeling the sphere of null directions at
the origin and $u_{ret}$ the retarded time. $\mathfrak{I}^{+}\ $is the limit
as $r$\ tends to infinity.

\begin{remark}
Note that $u_{ret},u,t\ $(and the variable $\tau \ $introduced later)$\ $all
have the dimensions of length. In the Section \ref{4}, the velocity of light, $c$, will be
explicitly introduced via the replacement ($u_{ret},u,t$, $\tau $) $\rightarrow(cu_{ret},cu,ct,c\tau)$ so that $u_{ret},u,t\ $and $\tau $
have the dimensions of time.
\end{remark}

\begin{remark}
We note that the round sphere metric, $ds^{2}=d\theta ^{2}+\sin ^{2}\theta
d\varphi ^{2}\ $becomes in stereographic coordinates $ds^{2}=4P^{-2}d\zeta d%
\overline{\zeta },\ $with $P=1+\zeta \overline{\zeta }.$
\end{remark}

To reach $\mathfrak{I}^{+}$, we simply let $r\rightarrow \infty $, so that $%
\mathfrak{I}^{+}$ has coordinates $(u,\zeta ,\bar{\zeta})$. The choice of a Bondi coordinate system is not unique,
there being a variety of Bondi coordinate systems to choose from. The
coordinate transformations between any two are know as Bondi-Metzner-Sachs
(BMS) transformations or as the BMS group (c.f., \cite{BMS1, BMS2}).

Our assumption of the analyticity of the space-time then allows for the
complexification of $\mathfrak{I}^{+}.$ For this complexification (i.e.,
extension to $\mathfrak{I}_{\mathbb{C}}^{+}$),$\ $we allow $u\ $to take on
complex values close to the real and free $\bar{\zeta}\ \ $from being the
complex conjugatae of $\zeta .\ $It is then denoted by$\ \widetilde{\zeta }%
\approx \bar{\zeta}.\ $(Often we take this as implicitly understood and just
use $\bar{\zeta}$.)

Associated with the Bondi coordinates is a (Bondi) null tetrad system, ($%
l^{a},n^{a},m^{a},\overline{m}^{a}$) (c.f., \cite{NT-80, NP:2009}):

\begin{eqnarray*}
l^{a}l_{a} &=&n^{a}n_{a}=m^{a}m_{a}=\overline{m}^{a}\overline{m}_{a}=0, \\
l^{a}n_{a} &=&-m^{a}\overline{m}_{a}=1.
\end{eqnarray*}%
\qquad \qquad The first tetrad vector $l^{a}$\ is the tangent to the
geodesics of the constant $u$ null surfaces given by

\begin{eqnarray}
l^{a}\ \  &=&\frac{\mathrm{d}x^{a}}{\mathrm{d}r}=g^{ab}\nabla _{b}u,
\label{Tet1} \\
l^{a}\nabla _{a}l^{b} &=&0,  \label{geo} \\
l^{a}\frac{\partial }{\partial x^{a}\ \ } &=&\frac{\partial }{\partial r}.
\label{Tet2}
\end{eqnarray}%
\ The second null vector $n^{a}$\ is tangent to the null geodesics lying on $%
\mathfrak{I}^{+}$, normalized so that: 
\begin{equation}
l_{a}n^{a}\ \ =1.  \label{Tet3}
\end{equation}%
The remaining vector $m^{a}\ $and its complex conjugate are tangent to the $%
S^{2}\ $slices of constant $u.\ $

An important construct is the family of past light-cones from each point of $%
\mathfrak{I}^{+}\ $(or $\mathfrak{I}_{\mathbb{C}}^{+}$).$\ $\ Each past cone
is determined by a sphere's worth of null directions at $\mathfrak{I}^{+}\ $%
with each null direction labeled by the associated sphere coordinate. \
These coordinates are chosen as the complex stereographic coordinates and
are denoted by the complex conjugate pair $(L,\overline{L})$. An arbitrary
field of null directions on $\mathfrak{I}^{+}\ $(and consequently an
arbitrary null geodesic congruence that intersects $\mathfrak{I}^{+}$) can
then be described by the function $L=L(u,\zeta ,\bar{\zeta})\ $or its
analytic extension to $\mathfrak{I}_{\mathbb{C}}^{+}.$

Often we will use a very specific form of a null tetrad given in Minkowski
coordinates and parametrized by the points on the sphere in stereographic
coordinates $(\zeta ,\overline{\zeta })\ $and denoted by the over-hat:%
\begin{eqnarray*}
\hat{l}^{a}\ \  &=&\frac{\sqrt{2}}{2(1+\zeta \bar{\zeta})}\left( 1+\zeta 
\bar{\zeta},\ \zeta +\bar{\zeta},\ -i(\zeta -i\bar{\zeta}),\ -1+\zeta \bar{%
\zeta}\right) =\left(\frac{\sqrt{2}}{2},\frac{1}{2}Y_{1i}^{(0)}\right), \\
\widehat{n}^{a}\ \  &=&\frac{\sqrt{2}}{2(1+\zeta \bar{\zeta})}\left( 1+\zeta 
\bar{\zeta},\ -(\zeta +\bar{\zeta}),\ i(\zeta -i\bar{\zeta}),\ 1-\zeta \bar{%
\zeta}\right) =\left(\frac{\sqrt{2}}{2},-\frac{1}{2}Y_{1i}^{(0)}\right), \\
\hat{m}^{a} &=&\eth l^{a}=\frac{\sqrt{2}}{2(1+\zeta \bar{\zeta})}(0,1-%
\overline{\zeta }^{2},-i(1+\overline{\zeta }^{2}),\text{ }2\overline{\zeta }%
)=(0,-Y_{1i}^{(1)}), \\
\overline{\hat{m}}^{a} &=&\overline{\eth }l^{a}=\frac{\sqrt{2}}{2(1+\zeta 
\bar{\zeta})}(0,1-\zeta ^{2},\text{ }i(1+\zeta ^{2}),2\zeta
),=(0,-Y_{1i}^{(-1)}).
\end{eqnarray*}

As $(\zeta ,\overline{\zeta })\ $move over the complex plane, $\hat{l}^{a}\ $%
and $\widehat{n}^{a}\ $range over the light cone. The spin-$s\ $harmonics \cite{NSO:2006}, $%
Y_{l,ijk....}^{(s)}(\zeta ,\bar{\zeta}),\ $which are frequently used, are
described in Appendix \ref{A}.

\subsection{Flat space Null Geodesic Congruences}

In Minkowski space $\mathbb{M}$,\ the future light-cones from an arbitrary
time-like world-line $x^{a}=\xi ^{a}(s)\ $can be described by the Null
Geodesic Congruence (NGC)%
\begin{equation}
x^{a}=\xi ^{a}(s)+r\hat{l}^{a}(\zeta ,\overline{\zeta })  \label{world-line}
\end{equation}%
with $r\ $the affine parameter on each of the light-cone generators. This
construct is easily generalized to complex Minkowski space $\mathbb{M}_{%
\mathbb{C}}$, where light-cones from $\xi ^{a}(\tau )$ (now an arbitrary
complex analytic world-line with complex affine parameter $\tau $) and its
corresponding \textit{complex} NGC is 
\begin{equation}
z^{a}=\xi ^{a}(\tau )+r\hat{l}^{a}(\zeta ,\widetilde{\zeta }),
\label{complex world-line}
\end{equation}%
where $r\ $is$\ $now complex and $(\zeta ,\widetilde{\zeta })\ $are
independent of each other.

The Sachs complex optical parameters for an arbitrary NGC (real or complex)
are the complex divergence and shear of the congruence \cite{NP:1962,NP:2009},

\begin{eqnarray}
\rho  &=&\frac{1}{2}(-\nabla _{a}l^{a}+i\text{ curl }l^{a}),  \label{rho} \\
\sigma  &=&\nabla _{(a}l_{b)}m^{a}m^{b}  \label{sigma1}
\end{eqnarray}%
where 
\begin{equation*}
\text{curl }l^{a}\equiv \sqrt{(\nabla _{\lbrack a}l_{b]}\nabla ^{a}l^{b})}.
\end{equation*}%
These satisfy the flat-space optical equations: 
\begin{eqnarray}
D\rho  &=&\rho ^{2}+\sigma \sigma   \label{optical} \\
D\sigma  &=&(\rho +\overline{\rho })\sigma ,  \notag \\
D &=&l^{a}\frac{\partial }{\partial x^{a}}=\frac{\partial }{\partial r}, 
\notag
\end{eqnarray}%
with $r\ $the affine parameter along the geodesics. The optical parameters
for the above light cone congruence can be calculated directly from Eq. (\ref%
{world-line}) yielding%
\begin{eqnarray}
\rho  &=&-r^{-1},  \label{light-cone} \\
\sigma  &=&0.  \notag
\end{eqnarray}

By reversing the statement and assuming a NGC with vanishing shear and real
divergence, the optical equations become%
\begin{equation*}
D\rho =\rho ^{2}.
\end{equation*}%
The integral (i.e., $\rho =-r^{-1}$) is the same as Eq.(\ref{light-cone}),
thus showing that a NGC with real divergence and vanishing shear is the
light-cone congruence of a (real) time-like world-line.

An arbitrary NGC in Minkowski space can be described by the three parameter, 
$(u,\zeta ,\overline{\zeta }),\ $family of null geodesics%
\begin{equation}
x^{a}=u(\hat{l}^{a}\ +\hat{n}^{a})-\overline{L}\hat{m}^{a}-L\overline{\hat{m}%
}^{a}+(r-r_{0})\hat{l}^{a}\   \label{NGC}
\end{equation}%
where $r\ $is the affine parameter, $r_{0}=r_{0}(u,\zeta ,\overline{\zeta }%
)\ $is the arbitrary origin for the affine parameter and $L=L(u,\zeta ,%
\overline{\zeta }),\ $the determining function of the congruence, is an
arbitrary complex spin-weight one function of the parameters. The three
parameters $(u,\zeta ,\overline{\zeta })\ \ $are the Bondi coordinates of
the intersection points of the null geodesics with $\mathfrak{I}^{+}$.$\ \ $%
The optical parameters are determined by $L\ $which is the stereographic
angle field on $\mathfrak{I}^{+}\ $that determines the directions of the
null geodesics$.$ \qquad \qquad

The condition for a NGC with vanishing shear is that the function $L\ $must
satisfy the non-linear partial differential equation \cite{Aronson:1972}
\begin{equation}
\eth L+L\dot{L}=0,  \label{shear=0}
\end{equation}%
where $\dot{L}=L,_{u}$, and $\eth$ is the spin-weighted covariant
derivative on the 2-sphere (see Appendix \ref{A} for details) \cite{Eth67}. \ This can be
integrated by introducing an auxiliary complex variable $\tau =T(u,\zeta ,%
\overline{\zeta }),\ $related to $L\ $by the so-called CR equation (related to the existence of a CR structure on $\mathfrak{I}^{+}$ \cite{NN:2006}),%
\begin{equation}
\eth T+L\dot{T}=0,  \label{CR}
\end{equation}%
and then using its inversion%
\begin{eqnarray}
u &=&G(\tau ,\zeta ,\overline{\zeta }),  \label{G} \\
\tau  &=&T(G(\tau ,\zeta ,\overline{\zeta }),\zeta ,\overline{\zeta })\equiv
\tau .
\end{eqnarray}%
After a process of implicit differentiation (c.f., \cite{HN75, Kozameh:2006zw, Adamo:2009vu}), Eq.(\ref{shear=0}) becomes%
\begin{equation}
\eth _{\tau }^{2}G=0,  \label{good cut eq.1}
\end{equation}%
with%
\begin{equation*}
L=\eth _{\tau }G|_{\tau =T(u,\zeta ,\overline{\zeta })},
\end{equation*}%
the subscript $\tau \ $indicates that the differentiation is at $\tau $\
held constant. \ From this it follows that the regular solutions to Eq.(\ref%
{shear=0}) can be given implicity in terms of $G$ as:

\begin{eqnarray}
u &=&G(\tau ,\zeta ,\overline{\zeta })\equiv \xi ^{a}(\tau )\hat{l}%
_{a}\Longleftrightarrow \tau =T(u,\zeta ,\overline{\zeta }),
\label{flatspace solution} \\
L &=&\eth G\equiv \xi ^{a}(\tau )\hat{m}_{a}=\xi ^{a}(T(u,\zeta ,\overline{%
\zeta }))\hat{m}_{a}.  \notag
\end{eqnarray}%
Several remarks must be made here:
\begin{itemize}
\item \ $\xi ^{a}(\tau )\ $are four arbitrary complex analytic functions of the
complex parameter $\tau \ $which can be interpreted as determining a complex$%
\ $world-line in complex$\ $Minkowski space.

\item $\tau $\ can be regauged by the$\ $analytic function $\tau ^{\ast
}=F(\tau ).\ $Often it is useful to chose $\xi ^{0}(\tau )\equiv \tau .$

\item Since $\tau \ $is complex we must allow $u\ $to take complex values which
requires the complexification of $\mathfrak{I}^{+}$, denoted $\mathfrak{I}_{%
\mathbb{C}}^{+}$.

\item \ When the $\xi ^{a}(\tau )\ $are real functions of a real variable, $s$,$%
\ $Eq.(\ref{NGC}) reduces to Eq.(\ref{world-line}) (i.e. to the real
world-line light cone congruence).

\item \ For a complex set $\xi ^{a}(\tau ),\ $the NGC, Eq.(\ref{NGC}), is a real
shear-free NGC but with a \textit{non-vanishing} twist. \ The caustic set is
in general a closed curve moving in time.\qquad 
\end{itemize}
An important observation that plays a major role for us is the following:
From the same $L,\ $two different \textquotedblleft
conjugate\textquotedblright\ versions can be constructed.\textit{\ }The
first is obviously the complex conjugate given by $\overline{L}=\overline{%
\xi }^{a}(\overline{\tau })\overline{\hat{m}}_{a}\ $while the second,
referred to as the holomorphic conjugate, is given by $\widetilde{L}=%
\overline{\eth }G=\xi ^{a}(\tau )\overline{\hat{m}}_{a}.\ $Using $\widetilde{%
L}\ $in Eq.(\ref{NGC}) instead of $\overline{L}$, we obtain another shear
free NGC but now it is the complex congruence, given earlier by Eq.(\ref%
{complex world-line}),%
\begin{equation*}
z^{a}=\xi ^{a}(\tau )+r\hat{l}^{a}(\zeta ,\widetilde{\zeta }).
\end{equation*}%

In other words, the cut function describes a family of null cones with apex on the complex line, $%
z^{a}=\xi ^{a}(\tau )$.\ \ We now have on $\mathfrak{I}^{+}$ and $\mathfrak{%
I}_{\mathbb{C}}^{+}\ $two different tetrad systems (obtained by null rotations from the
Bondi tetrad) coming from $\overline{L}\ $and $\widetilde{L}$, namely
\begin{eqnarray}
l^{a} &\rightarrow &l^{\ast a}=l^{a}-\frac{\overline{L}}{r}m^{a}-\frac{L}{r}%
\overline{m}^{a}+O(r^{-2}),  \label{null.rotI} \\
m^{\ast a} &=&m^{a}-\frac{\overline{L}}{r}n^{a}, \\
n^{\ast a} &=&n^{a}
\end{eqnarray}%
and%
\begin{eqnarray}
l^{a} &\rightarrow &l_{C}^{\,\ast a~}=l^{a}-\frac{\widetilde{L}}{r}m^{a}-%
\frac{L}{r}\overline{m}^{a}+O(r^{-2}),\   \label{null.rotII} \\
m^{\ast a} &=&m^{a}-\frac{\widetilde{L}}{r}n^{a}, \\
n^{\ast a} &=&n^{a}.
\end{eqnarray}%
The null geodesic congruence determined by $l^{\ast a}$, as mentioned
earlier, is a real shear-free congruence with twist while the congruence
determined by $l_{C}^{\,\ast a~}$is complex, shear-free, twist free
conguence and focuses on the complex curve $\xi ^{a}(\tau ).$

Though the complex null geodesics with apex on $\xi ^{a}(\tau )\ $spend most
of their \textquotedblleft time\textquotedblright\ in the complex Minkowski
space some do reach real Minkowski space and in particular some reach the
real $\mathfrak{I}^{+}.\ $It turns out that the complex world-line and their
associated light-cones have real structures. They are discussed in Section \ref{3}.

\subsection{Asymptotic Flatness}

At a first glance it would appear as if it were not possible to duplicate
the Minkowski space discussion of light-cone NGCs in asymptotically flat
space-times. Aside for a few special cases (the algebraically special
metrics) there are no Einstein space-times with shear-free NGCs. The family
of future directed null geodesics originating at a fixed space-time point
traversing regions of curvature will, in general, be distorted and develop
shear. Surprisingly it nevertheless is possible to duplicate virtually all
the light-cone NGC results of flat-spacetime for the general case of
asymptotically flat space-times by looking not for shear-free NGCs but
instead \textit{asymptotically shear-free conguences.} In fact such
congruences are determined by a complex analytic curve in an auxiliary
four-complex dimensional space, referred to as $\mathcal{H}$-space.

Before describing these congruences we first review some relevant features
of asympotically flat space-times. Details, derivations and proofs are
largely omitted since they are easily found in the literature \cite{NP:1962, NP:2009, NT-80, Adamo:2009vu}.

We begin by pointing out that with Bondi coordinates and tetrad the two
optical parameters, the complex divergence and shear are given by%
\begin{eqnarray}
\rho  &=&\overline{\rho }=-\frac{1}{r}+\frac{\sigma ^{0}\overline{\sigma }%
^{0}}{r^{3}}+O(r^{-5}),  \label{div&shear} \\
\sigma  &=&\frac{\sigma ^{0}}{r^{2}}+O(r^{-4}),  \notag
\end{eqnarray}%
with $\sigma ^{0}=\sigma ^{0}(u,\zeta ,\overline{\zeta }),$ the asymptotic
Bondi shear of the NCG with the Bondi tangent vector, i.e., $l^{a}.\ $The $\sigma
^{0}$, which is the free data determining the gravitational radiation,
plays a major role in \ our discussion. Considering a new NGC with tangent
vector \ $l^{\ast a}\ $defined at $\mathfrak{I}^{+}\ $by the null rotation

\begin{eqnarray}
l^{\ast a} &=&l^{a}-\frac{\overline{L}}{r}m^{a}-\frac{L}{r}\overline{m}%
^{a}+O(r^{-2}),  \label{null rot2} \\
m^{\ast a} &=&m^{a}-\frac{\overline{L}}{r}n^{a}, \\
n^{\ast a} &=&n^{a}
\end{eqnarray}%
with arbitrary $L=L(u,\zeta ,\overline{\zeta }),\ $one finds that the
asymptotic shear of the new conguence is given by a version of the Sachs theorem \cite{Aronson:1972}:%
\begin{equation}
\sigma ^{0\ast }=\eth L+L\dot{L}-\sigma ^{0}.  \label{shear*}
\end{equation}%
The condition for the new congruences to be asymptotically shear-free ($%
\sigma ^{0\ast }=0$) is thus that $L\ $satisfy%
\begin{equation}
\eth L+L\dot{L}=\sigma ^{0},  \label{shear=0*}
\end{equation}%
which is the extension of the flat-space Eq.(\ref{shear=0}).

As in the Minkowski space case, Eq.(\ref{shear=0}), this can also be
integrated by introducing the same auxiliary complex variable $\tau
=T(u,\zeta ,\overline{\zeta }),\ $related to $L\ $by the CR equation,%
\begin{equation}
\eth T+L\dot{T}=0,  \label{CR*}
\end{equation}%
and then using its inversion%
\begin{eqnarray}
u &=&G(\tau ,\zeta ,\overline{\zeta }),  \label{complex cuts} \\
\tau  &=&T(G(\tau ,\zeta ,\overline{\zeta }),\zeta ,\overline{\zeta })\equiv
\tau .  \label{and inverse}
\end{eqnarray}

Note that as in the flat case, $\tau \ $is complex$\ $and we must allow the
complexification of $u\ $and let $\widetilde{\zeta }\approx \overline{\zeta }%
.\ $For each value of $\tau $ we obtain a complex \textquotedblleft
cut\textquotedblright\ of $\mathfrak{I}_{\mathbb{C}}^{+}$.

Again after manipulating several implicit derivatives, Eq.(\ref{shear=0*})
and (\ref{CR*}) become%
\begin{eqnarray}
\eth _{\tau }^{2}G &=&\sigma ^{0}(G,\zeta ,\overline{\zeta }),  \label{GCEq}
\\
L(u,\zeta ,\overline{\zeta }) &=&\eth _{\tau }G|_{\tau =T(u,\zeta ,\overline{%
\zeta })}  \label{L*}
\end{eqnarray}%
with again the subscript $\tau \ $indicating that the derivatives are at $%
\tau $\ held constant. \ Eq.(\ref{GCEq}), the \textquotedblleft good cut
equation,\textquotedblright\ has been shown to depend on four complex
parameters, $z^{a},\ $(the $\mathcal{H}$-space coordinates), so that we can
write, $u=X(z^{a},\zeta ,\overline{\zeta })$, where we distinguish $X$ from $G$ by its explicit dependence on the four solution parameters.  By the coordinate freedom, 
\begin{equation*}
z^{a}\rightarrow z^{\ast a}=f^{a}(z^{a}),
\end{equation*}%
the first four spherical harmonic coefficients can be chosen as the $z^{a}$,
(coordinate conditions on $\mathcal{H}$-space)$\ $so that we have 
\begin{equation}
u=X(z^{a},\zeta ,\overline{\zeta })=z^{a}\hat{l}_{a}(\zeta ,\overline{\zeta }%
)+H_{l\geqslant 2}(z^{a},\zeta ,\overline{\zeta }),  \label{solution II}
\end{equation}%
where $H_{l\geq 2}$ are spherical harmonic contributions with $l\geq 2$.
Finally, by taking an arbitrary world-line in the $\mathcal{H}$-space, $%
z^{a}=\xi ^{a}(\tau ),\ $we find the general regular solution to Eq.(\ref%
{shear=0*}) is given implicitly by%
\begin{eqnarray}
L(u,\zeta ,\overline{\zeta }) &=&\eth _{\tau }G|_{\tau =T(u,\zeta ,\overline{%
\zeta })}  \label{solution*} \\
G(\tau ,\zeta ,\overline{\zeta }) &\equiv &X(\xi ^{a}(\tau ),\zeta ,%
\overline{\zeta })  \label{X} \\
u &=&G(\tau ,\zeta ,\overline{\zeta })  \label{u=X} \\
&=&\xi ^{a}(\tau )\hat{l}_{a}(\zeta ,\overline{\zeta })+H_{l\geqslant 2}(\xi
^{a}(\tau ),\zeta ,\overline{\zeta })=\xi ^{a}(\tau )\hat{l}_{a}(\zeta ,%
\overline{\zeta })+\tilde{H}_{l\geqslant 2}(\tau ,\zeta ,\overline{\zeta }).
\notag
\end{eqnarray}

As in the flat space case, the regular asymptotically shear-free NGCs are
determined by the arbitrary choice of a complex world-line in an auxiliary
complex space, $\mathcal{H}$-space.

In complete analogy with the complex Minkowski space case, it turns out that
if we use the complex NGC determined by the stereographic angle field, (\ref%
{solution*}), and the associated holomorphic field, $\tilde{L}(u,\zeta ,%
\overline{\zeta })=\overline{\eth }_{\tau }G|_{\tau =T(u,\zeta ,\overline{%
\zeta })},\ $(not $\overline{L}(u,\zeta ,\overline{\zeta })=\overline{\eth }%
_{\tau }\overline{G}|_{\overline{\tau }=\overline{T}(u,\zeta ,\overline{%
\zeta })}$) as initial directions at $\mathfrak{I}_{\mathbb{C}}^{+}$%
\begin{equation}
l_{C}^{\ast a}=l^{a}-\frac{\tilde{L}}{r}m^{a}-\frac{L}{r}\overline{m}%
^{a}+O(r^{-2}),  \label{complex l}
\end{equation}%
the complex geodesics converge on the $\mathcal{H}$-space world-line $%
z^{a}=\xi ^{a}(\tau )$ \cite{Ludvigsen:1980}. By pointing into complex null directions from the
complexified $\mathfrak{I}_{\mathbb{C}}^{+}$,\ we have complex \textit{%
virtual} \textit{cones} and complex \textit{virtual world-lines}. There will
always be points on the complex world-line whose null cones partially
intersect real $\mathfrak{I}^{+}$. We will see later that unique world-lines
can be determined so that \textit{real} meaning or significance can be given
to them, as complex centers of charge and complex centers of mass - which
include both asymptotic magnetic dipoles and angular momentum.

The basic idea that will be pursued later (in Sections \ref{3} and \ref{4}) is to
identify certain terms in the asymptotic behavior of the Maxwell and Weyl tensor tetrad
components with physical quantities and then see how they change when they
are computed with the (rotated) complex null directions pointing towards a
complex world-line, Eq.(\ref{complex l}). \ By choosing the world-line
appropriately, so that these quantities vanish, we identify the \textit{%
virtual} complex centers of charge and mass.

As an interim step, we need the behavior of both the tetrad components of
the Weyl and Maxwell tensors:

\begin{eqnarray}
\psi _{0} &=&-C_{abcd}l^{a}m^{b}l^{c}m^{d},  \label{weyl} \\
\psi _{1} &=&-C_{abcd}l^{a}n^{b}l^{c}m^{d},  \notag \\
\psi _{2} &=&-\frac{1}{2}%
(C_{abcd}l^{a}n^{b}l^{c}n^{d}-C_{abcd}l^{a}n^{b}m^{c}\overline{m}^{d}), 
\notag \\
\psi _{3} &=&C_{abcd}l^{a}n^{b}n^{c}\overline{m}^{d},  \notag \\
\psi _{4} &=&C_{abcd}n^{a}\overline{m}^{b}n^{c}\overline{m}^{d},  \notag
\end{eqnarray}%
and 
\begin{align}
\phi _{0}& =F_{ab}l^{a}m^{b},  \label{maxwell} \\
\phi _{1}& =\frac{1}{2}F_{ab}(l^{a}n^{b}+m^{a}\overline{m}^{b}),  \notag \\
\phi _{2}& =F_{ab}n^{a}\overline{m}^{b}.  \notag
\end{align}

Integrating the both the Weyl tensor and Maxwell spin-coefficient equations
leads to the peeling behavior:

\begin{eqnarray}
\psi _{0} &=&\psi _{0}^{0}r^{-5}+O(r^{-6}),  \label{peeling} \\
\psi _{1} &=&\psi _{1}^{0}r^{-4}+O(r^{-5}),  \notag \\
\psi _{2} &=&\psi _{2}^{0}r^{-3}+O(r^{-4}),  \notag \\
\psi _{3} &=&\psi _{3}^{0}r^{-2}+O(r^{-3}),  \notag \\
\psi _{4} &=&\psi _{4}^{0}r^{-1}+O(r^{-2}),  \notag
\end{eqnarray}%
and 
\begin{eqnarray}
\phi _{0} &=&\phi _{0}^{0}r^{-3}+O(r^{-4}),  \label{peeling II} \\
\phi _{1} &=&\phi _{1}^{0}r^{-2}+O(r^{-3}),  \notag \\
\phi _{2} &=&\phi _{2}^{0}r^{-1}+O(r^{-2}).  \notag
\end{eqnarray}%
With the coefficients satisfying the asymptotic Bianchi Identities%
\begin{eqnarray}
\dot{\psi}_{2}^{0\,} &=&-\eth \psi _{3}^{0\,}+\sigma ^{0\,}\psi^{0} _{4},
\label{AsyBI1*} \\
\dot{\psi}_{1}^{0\,} &=&-\eth \psi _{2}^{0\,}+2\sigma ^{0}\psi _{3}^{0\,},
\label{AsyBI2*} \\
\dot{\psi}_{0}^{0\,} &=&-\eth \psi _{1}^{0\,}+3\sigma ^{0}\psi _{2}^{0\,},
\label{AsyBI3*}
\end{eqnarray}%
where
\begin{eqnarray}
\psi _{4}^{0} &=&-\ddot{\overline{\sigma }}^{0},  \label{i} \\
\psi _{3}^{0} &=&\eth \dot{\overline{\sigma }}^{0},  \label{j}
\end{eqnarray}%
and asymptotic Maxwell equations
\begin{eqnarray}
\dot{\phi}_{1}^{0} &=&-\eth \phi _{2}^{0},  \label{AsyBI4} \\
\dot{\phi}_{0}^{0} &=&-\eth \phi _{1}^{0}+\sigma ^{0}\phi _{2}^{0}.
\label{AsyBI5}
\end{eqnarray}

All the $r^{-n}\ $coefficients, which are functions on $\mathfrak{I}^{+}\ $,
i.e., functions of $(u,\zeta ,\overline{\zeta })$,$\ $have physical meaning,
e.g., multipole moments, etc. The four quantities \{$\psi _{1}^{0},\psi
_{2}^{0},\phi _{0}^{0},\phi _{1}^{0}$\} are the most importance to us, due
to the following properties:
\begin{itemize}
\item The $l=0\ $harmonic of $\phi _{1}^{0}\ $is the Coulomb charge $%
q.\ $It is assumed to be nonvanishing whenever a Maxwell field is being
considered.

\item The $l=1\ $harmonic of $\phi _{0}^{0}\ $is the complex
electromagnetic dipole moment: $D_{E\&M}^{i}=D_{E}^{i}+iD_{M}^{i}$.

\item The $l=0,1\ $harmonics of $\psi _{2}^{0},\ $slightly modified by
the shear$\ \sigma ^{0}$,$\ $is the Bondi energy-momentum four-vector.

\item The $l=1\ $harmonic of $\psi _{1}^{0},\ $also slightly modified$%
\ $by the shear$\ \sigma ^{0}$,$\ $encodes the center of mass dipole and
angular momentum: $D_{\mathbb{C}(grav)}^{i}=D_{\text{mass}}^{i}+ic^{-1}J^{i}$.
\end{itemize}

By using the tetrad transformation generated by Eq.(\ref{complex l}) (see
Eq.(\ref{null.rotII})), one find the transformation law of the leading terms
of the Weyl and Maxwell tensors: 
\begin{eqnarray}
\psi _{0}^{\ast 0} &=&\psi _{0}^{0}-4L\psi _{1}^{\ast 0}+6L^{2}\psi
_{2}^{\ast 0}-4L^{3}\psi _{3}^{\ast 0}+L^{4}\psi _{4}^{0},  \label{Weyl rot}
\\
\psi _{1}^{\ast 0} &=&\psi _{1}^{0}-3L\psi _{2}^{0}+3L^{2}\psi
_{3}^{0}-L^{3}\psi _{4}^{0},  \label{Weyl rot2} \\
\psi _{2}^{\ast 0} &=&\psi _{2}^{0}-2L\psi _{3}^{0}+L^{2}\psi _{4}^{0},
\label{Weyl rot3} \\
\psi _{3}^{\ast 0} &=&\psi _{3}^{0}-L\psi _{4}^{0},  \label{Weyl rot4} \\
\psi _{4}^{0} &=&\psi _{4}^{\ast 0},  \label{Weyl rot5}
\end{eqnarray}%
\begin{eqnarray}
\phi _{0}^{\ast 0} &=&\phi _{0}^{0}-2L\phi _{1}^{0}+L^{2}\;\phi _{2}^{0},
\label{transformed} \\
\phi _{1}^{\ast 0} &=&\phi _{1}^{0}-L\phi _{2}^{0},  \label{transformed*} \\
\phi _{2}^{\ast 0} &=&\phi _{2}^{0}.  \label{transformed**}
\end{eqnarray}

Later, setting to zero the $l=1\ $parts of $\psi _{1}^{\ast 0}\ $and $\phi
_{0}^{\ast 0},\ $we can determine two different world-lines (when a Maxwell
field is present) that can be referred to respectively as the complex centers of mass and
charge.

\section{Real Structures from the Complex World-line}
\label{3}

Our task in this section is to find the real structures that are lying in
the complex world-lines and their complex light-cones.

\subsection{Flat-space Real Structure}

We first examine the case of flat-space-time with a complex Minkowski
world-line, $z^{a}=\xi ^{a}(\tau )\ $and associated light-cone cut of $%
\mathfrak{I}_{\mathbb{C}}^{+},\ $%
\begin{equation}
u=\xi ^{a}(\tau )l_{a}(\zeta ,\overline{\zeta }).  \label{l.c.cut}
\end{equation}%
To answer our question: what values of $\tau \ $allow real values of $u,\ $%
we first write $\tau =s+i\lambda ,\ (s,\lambda \ $real$)\ $and decompose the
right-hand side of Eq.(\ref{l.c.cut}) into its real and imaginary parts and
set the imaginary part to zero \cite{Adamo:2009ru}, 
\begin{eqnarray}
u &=&\frac{1}{2}(\xi ^{a}(s+i\lambda )l_{a}(\zeta ,\overline{\zeta })+%
\overline{\xi }^{a}(s-i\lambda )l_{a}(\zeta ,\overline{\zeta }))
\label{real&imag} \\
&&+\frac{1}{2}(\xi ^{a}(s+i\lambda )l_{a}(\zeta ,\overline{\zeta })-%
\overline{\xi }^{a}(s-i\lambda )l_{a}(\zeta ,\overline{\zeta }),  \notag \\
0 &=&[\xi ^{a}(s+i\lambda )-\overline{\xi }^{a}(s-i\lambda )]l_{a}(\zeta ,%
\overline{\zeta }).  \label{imag}
\end{eqnarray}%
Considering Eq.(\ref{imag}) as an implicit equation defining 
\begin{equation}
\lambda =\Lambda (s,\zeta ,\overline{\zeta })  \label{Lambda}
\end{equation}%
we have that the allowed values of $\tau \ $are given by 
\begin{equation}
\tau =s+i\Lambda (s,\zeta ,\overline{\zeta }).  \label{allowed tau}
\end{equation}

The real values of $u\ $are thus given by the one-parameter ($s$)\ family of
slicings%
\begin{equation}
u=\xi ^{a}(s+i\Lambda (s,\zeta ,\overline{\zeta }))%
l_{a}(\zeta ,\overline{\zeta }).  \label{real u}
\end{equation}

Assuming small values for the imaginary part of $\xi ^{a}(\tau )=\xi
_{R}^{a}(\tau )+i\xi _{I}^{a}(\tau )$, $(\xi _{R}^{a}(\tau ),\ \xi
_{I}^{a}(\tau ))$ both real analytic functions) and hence small $\Lambda
(s,\zeta ,\overline{\zeta })$, it has been shown that $\Lambda (s,\zeta ,%
\overline{\zeta }),\ $(for fixed value of $s$), is a bounded smooth function
on the $(\zeta ,\overline{\zeta })\ $sphere, with maximum and minimum
values, $\lambda _{\max }=\Lambda (s,\zeta _{\max },\overline{\zeta }_{\max
})\ $and $\lambda _{\min }=\Lambda (s,\zeta _{\min },\overline{\zeta }_{\min
}).\ $Furthermore on the sphere, there are a circle's ($S^{1}$) worth of
curves between $(\zeta _{\min },\overline{\zeta }_{\min })\ $and $(\zeta
_{\max },\overline{\zeta }_{\max })\ $such that $\Lambda (s,\zeta ,\overline{%
\zeta })\ $is a monotonically increasing function on each curve. Hence$\ $%
there will be a family of circles on the $(\zeta ,\overline{\zeta })$-sphere
where value of $\lambda \ $is a constant, ranging between $\lambda _{\max }\ 
$and $\lambda _{\min }$.

Summarizing, we have the result that in the complex $\tau $-plane there is a
ribbon or strip given by all values of $s\ $and line segment parameterized
by $\lambda \ $between $\lambda _{\min }\ $and $\lambda _{\max }\ $such that
the complex light-cones from each of the associated points, $\xi
^{a}(s+i\lambda ),\ $all have \textit{some null geodesics} that intersect
real $\mathfrak{I}^{+}$.\ \ More specifically, for each allowed value of $%
\tau =s+i\Lambda \ $there will be a circle's worth of complex null geodesics
leaving the point $\xi ^{a}(s+i\lambda )\ $reaching real $\mathfrak{I}^{+}.\ 
$It is the union of these null geodesics, corresponding to the circles on
the $(\zeta ,\overline{\zeta })$-sphere from the line segment, that produces
the real family of cuts, Eq.(\ref{real u}).

The real structure associated with a complex world-line is then the
one-pameter family of slices (cuts) Eq.(\ref{real u}) and angle field $L(u, \zeta,\bar{\zeta})$ on each point of the cuts.

The dual point of view, as previously mentioned, is to start with the same $%
L $ as used earlier:

\begin{eqnarray}
u &\equiv &\xi ^{a}(\tau )\hat{l}_{a}\Longleftrightarrow \tau =T(u,\zeta ,%
\overline{\zeta }),  \label{flatspace sol.II} \\
L &=&\overline{\eth }_{\tau }G\equiv \xi ^{a}(\tau )\hat{m}_{a}=\xi
^{a}(T(u,\zeta ,\overline{\zeta }))\hat{m}_{a}.  \notag
\end{eqnarray}%
which was used with the holomorphic $\widetilde{L},$ 
\begin{equation*}
\widetilde{L}=\eth _{\tau }G\equiv \xi ^{a}(\tau )\overline{\hat{m}}_{a}=\xi
^{a}(T(u,\zeta ,\overline{\zeta }))\overline{\hat{m}}_{a},
\end{equation*}%
but now, instead, use the complex conjugate of $L$: 
\begin{equation*}
\overline{L}=\overline{\eth }_{\tau }\overline{G}=\overline{\xi }^{a}(%
\overline{\tau })\overline{\hat{m}}_{a}
\end{equation*}%
for the null directions pointing inward. \ In this case one obtains again a real
shear-free NGC but now with twist $\Sigma (u,\zeta ,\overline{\zeta })\ $%
which comes from the complex divergence,%
\begin{eqnarray}
\rho  &=&-\frac{1}{r+i\Sigma }  \label{complex div} \\
2i\Sigma  &=&\eth \overline{L}+L(\overline{L})^{\cdot }-\overline{\eth }L-%
\overline{L}\dot{L}.  \label{twist} \\
&=&(\xi ^{a}(\tau )-\overline{\xi }^{a}(\overline{\tau }))\left(
n_{a}-l_{a}\right) .  \notag
\end{eqnarray}%
As was claimed earlier, the twist is proportional to the imaginary part of
the complex world-line and consequently we have the real structure coming
from two (dual) places.

\subsection{Asymptotically Flat-Space Real Structure}

The extension of the above argument to the case of asymptotically flat
space-times is relatively simple. \ Again assuming that the Bondi shear is
sufficiently small and the $\mathcal{H}$-space complex world-line is not too
far from the \textquotedblleft real,\textquotedblright\ the solution to the
good-cut equation (\ref{GCEq}), 
\begin{equation}
u=\xi ^{a}(\tau )\hat{l}_{a}(\zeta ,\overline{\zeta })+\tilde{H}_{l\geqslant
2}(\tau ,\zeta ,\overline{\zeta })\equiv G(\tau ,\zeta ,\overline{%
\zeta }),  \label{solution 2}
\end{equation}%
with $\tau =s+i\lambda ,\ $is decomposed into real and imaginary parts,%
\begin{equation}
G(\tau ,\zeta ,\overline{\zeta })=\frac{1}{2}{\large (}G(s+i\lambda ,\zeta ,%
\overline{\zeta })+\overline{G}(s-i\lambda ,\zeta ,\overline{\zeta }){\large %
)}+\frac{1}{2}{\large (}G(s+i\lambda ,\zeta ,\overline{\zeta })-\overline{G}%
(s-i\lambda ,\zeta ,\overline{\zeta }){\large ).}  \label{decomposition}
\end{equation}%
\qquad 

Setting the imaginary part to zero and solving for $\lambda \ $we obtain an
expression of the form,%
\begin{equation*}
\lambda =\Lambda (s,\zeta ,\overline{\zeta }).
\end{equation*}%
As in the flat case, for fixed $s=s_{0}$, $\Lambda \ $has values on a line
segment bounded between some $\lambda _{\min \ }$and $\lambda _{\max \ }$.
The allowed values of $\tau \ $are again on a ribbon in the $\tau $-plane,
all values of $s\ $and values on the $\lambda $-line segments.

Each level curve of the function $\lambda =\Lambda (s_{0},\zeta ,\overline{%
\zeta })=$constant$\ $on the $(\zeta ,\overline{\zeta })$-sphere,$\ $(closed
curves or isolated points), determines a specific subset of the null
directions and associated null geodesics on the light-cone of the complex
point $\xi ^{a}(s_{0}+i\Lambda (s_{0},\zeta ,\overline{\zeta }))\ $that
intersect the real $\mathfrak{I}^{+}$.$\ $These geodesics will be referred
to as \textquotedblleft \textit{real\textquotedblright geodesics}. As $%
\lambda $\ moves over all allowed values of its segment, we obtain the$\ $%
set of $\mathcal{H}$-space points, $\xi ^{a}(s_{0}+i\Lambda (s_{0},\zeta ,%
\overline{\zeta }))$\ and their collection of \textquotedblleft
real\textquotedblright\ geodesics. \ From Eq.(\ref{decomposition}), these
\textquotedblleft real\textquotedblright\ geodesics\textit{\ }intersects $%
\mathfrak{I}^{+}\ $on the cut%
\begin{equation*}
u\ {\large =}G(s_{0}+i\Lambda (s_{0},\zeta ,\overline{\zeta }),\zeta ,%
\overline{\zeta }).
\end{equation*}

As $s\ $varies we obtain a one-parameter family of cuts. If these cuts do
not intersect with each other we say that the complex world-line $\xi
^{a}(\tau )\ $is by definition a \textquotedblleft
time-like\textquotedblright\ line. This occurs when the time component of
the real part of the complex velocity vector, $v^{a}(\tau )=\mathrm{d}\xi
^{a}(\tau )/\mathrm{d}\tau $,$\ $is sufficiently large.

\subsection{Summary of Real Structures}

To put the ideas of this section into perspective we collect the claims.

\begin{itemize}
\item In Minkowski space, the future directed light-cones emanating from a
real time-like world-line, $x^{a}=\xi ^{a}(s)$, intersect future null
infinity, $\mathfrak{I}^{+}$,\ on a one-parameter family of spherical
non-intersecting cuts.

\item The complex light-cones emanating from a time-like complex analytic
curve in complex Minkowski space,$\ z^{a}=\xi ^{a}(\tau )$ parametrized by
the complex parameter $\tau =s+i\lambda $,$\ $has for each fixed value of $%
s\ $and $\lambda \ $a limited set null geodesics that reach real $\mathfrak{I%
}^{+}.\ \ $However, for a ribbon in the complex $\tau $-plane (i.e., a
region topologically $\mathbb{R}\times I$, with $s\in \mathbb{R}$ and $%
\lambda \in I=[\lambda _{\min },\lambda _{\max }]$), there will be many null
geodesics intersecting $\mathfrak{I}^{+}.\ $Such null geodesics were
referred to as \textquotedblleft \textit{real\textquotedblright geodesics}
. \ More specifically, for a fixed $s$, there is a limited range of $\lambda
\ $such that all the real null geodesics intersect $\mathfrak{I}^{+}\ $in a
full cut, leading to a one-parameter family of real (distorted sphere)
slicings. \ The ribbon is the generalization of the real world-line and the
slicings are the analogues of the spherical slicings. When the ribbon
shrinks to a line it degenerates to the real case. We can consider the
ribbon as a generalized world-line and the \textquotedblleft
real\textquotedblright\ null geodesics\textit{\ }from constant $s\ $portion
of the ribbon as a generalized light-cone.

\item For the case of asymptotically flat space-times, the real light-cones
from interior points are replaced by the virtual light-cones generated by
the asymptotically shear-free NGCs. \ These cones emanate from a\ complex
virtual world-line $z^{a}=\xi ^{a}(\tau )\ $in the associated $\mathcal{H}$%
-space. \ As in the case of complex Minkowski space, there is a ribbon in
the $\tau $-plane where the \textquotedblleft real\textquotedblright\ null
geodesics\textit{\ }emanate from. \ The \textquotedblleft
real\textquotedblright\ null geodesics\textit{\ }coming from a cross-section
of the strip at fixed $s\ $(as in the complex Minkowski case), intersect $%
\mathfrak{I}^{+}\ $in a cut; the collection of cuts yielding a one-parameter
family of cuts. \ The situation is exactly the same as in the complex
Minkowski space case except that the spherical harmonic decomposition of
these cuts is in general more complicated.
\end{itemize}

\subsubsection*{\textit{Example: The (charged) Kerr metric}}
Considering the Kerr or the charged Kerr metrics (or even more generally any
asymptotically flat stationary metric), we have immediately that the Bondi
shear $\sigma ^{0}\ $vanishes and hence the associated $\mathcal{H}$-space
is complex Minkowski space \cite{Adamo:2009fq, Adamo:2010ey}. \ From the stationarity and a real origin shift
and rotation, the complex world-line can be put into the form%
\begin{equation}
\xi ^{a}(\tau )=(\tau ,0,0,ia),  \label{Kerr world-line}
\end{equation}%
with $a$ being the Kerr parameter. The complex cut function is then
\begin{eqnarray}
u\ \ &=&\xi ^{a}(\tau )\hat{l}_{a}(\zeta ,\bar{\zeta})  \label{Kerr cut} \\
&=&\frac{\tau }{\sqrt{2}}-\frac{i}{2}aY_{1,3}^{0}(\zeta ,\overline{\zeta }),
\notag \\
Y_{1,3}^{0}(\zeta ,\overline{\zeta }) &=&-\sqrt{2}\frac{1-\zeta \overline{%
\zeta }}{1+\zeta \overline{\zeta }},  \notag
\end{eqnarray}%
so that the angle fields are 
\begin{eqnarray*}
L &=&\sqrt{2}ia\frac{\overline{\zeta }}{1+\zeta \overline{\zeta }}, \\
\overline{L} &=&-\sqrt{2}ia\frac{\zeta }{1+\zeta \overline{\zeta }}, \\
\widetilde{L} &=&\sqrt{2}ia\frac{\zeta }{1+\zeta \overline{\zeta }}.
\end{eqnarray*}

\ Using $\tau =s+i\lambda \ $in Eq.(\ref{Kerr cut}), the reality condition $%
u=\overline{u}$ on the cut function is that%
\begin{equation*}
\lambda =\Lambda (s,\zeta ,\bar{\zeta})=\frac{\sqrt{2}}{2}aY_{1,3}^{0}(\zeta
,\overline{\zeta }),
\end{equation*}%
so that on the $\tau $-ribbon, $\lambda \ $ranges between $\pm $ $\sqrt{2}\ $%
and the real slices from the ribbon becomes simply $u=s/\sqrt{2}.$     $\Box$

Though we are certainly not making the claim that one can in reality
\textquotedblleft observe\textquotedblright\ these complex worle-lines that
arise from (asymptotically) shear-free congruences, we nevertheless claim
that they can be observed in a different sense. In the following section we
will show that there are simple physical measurements that do determine
these compex world-lines.

\section{Applications}
\label{4}

We can now explore uses of our observations concerning light-cones and their
generalizations. The first issue addressed is the application, in Minkowski
space, to the Maxwell equations and in particular to asymptotically
vanishing Maxwell fields with non-vanishing charge $q$. Specifically, we
show that such solutions naturally define a complex world-line that can be
identified or referred to as the complex center of charge. It is determined
from the complex Minkowski space points where the (suitably defined) \textit{%
complex electromagnetic dipole} (a combination of the electric dipole moment
plus \textquotedblleft $i$\textquotedblright\ magnetic dipole moment)
vanishes.

The analogous problem for asymptotically flat space-times (either vacuum or
Einstein-Maxwell) is addressed with a unique world-line again arising, this
time from the gravitational part with its identification as the complex
center of mass. These are the $\mathcal{H}$-space points where the \textit{%
complex gravitational dipole} (identified as the mass dipole plus
\textquotedblleft $i$\textquotedblright\ angular momentum) vanishes. For the
Einstein-Maxwell case there will be, in addition, a complex center of charge
line.

A few words of explanation in a much simpler situation might be of use. In
Minkowski space, in a given Lorentz frame and coordinate origin, with given
charge and current distributions (or given mass and spin distribution), one
defines the electric dipole moment (mass dipole) on any time slice by an
space integral over the charge density (or mass density) times the position.
\ By shifting the spatial origin, the dipole moment becomes a \textit{%
space-time field} depending on the origin shift, 
\begin{equation*}
\overrightarrow{D^{\ast }}=\overrightarrow{D}-q\overrightarrow{R}.
\end{equation*}%
The zero values of this field determine the center of charge (or center of
mass); $\overrightarrow{R}=q^{-1}\overrightarrow{D}$.

By extending this idea to include the magnetic dipole moment 
\begin{equation*}
\overrightarrow{D}_{E\& M}=\overrightarrow{D}_{E}+i\overrightarrow{D}%
_{M}.
\end{equation*}%
and allowing the position, $\overrightarrow{R}$, to take on complex values,
we find the space dependence of the complex dipole moments given by%
\begin{equation}
\overrightarrow{D}_{E\& M}^{\ast }=\overrightarrow{D}_{E\& M}-q%
\overrightarrow{R}_{\mathbb{C}}.  \label{D_C}
\end{equation}%
so that the complex center of charge is given by $\overrightarrow{D}_{E\& M}^{\ast }=0$ or 
\begin{equation}
\overrightarrow{R}_{\mathbb{C}}=q^{-1}\overrightarrow{D}_{E\& M}.
\end{equation}

The difficulty with this construction is that it is not Lorentz invariant:
the transformations of the dipoles from one Lorentz frame to another is
non-local and one does not obtain (in any obvious manner) a unique center of
charge/mass world-line.

We use an alternate procedure to find the different \textquotedblleft
centers of motion.\textquotedblleft\ Namely, the complex dipoles\ are first
identified from the asymptotic solutions with interior sources: they are
identified from the $l=1\ $harmonics in the tetrad components
(spin-coefficient components) of the asymptotic Maxwell field and the
asymptotic Weyl tensor, Eqs.(\ref{weyl}) and (\ref{maxwell}).\ (See the
discussion immediately after Eq.(\ref{peeling II}).) These quantities 
\textit{depend on the choice of the tetrad vectors} at $\mathfrak{I}^{+}.\ $%
If we choose the tetrad so that the null vector $l=l_{C}^{\ast }\ $%
determines a shear-free (or asymptotically shear-free) null geodesic
congruence that focuses on points in complex Minkowski space (or the GR
case, on points in the virtual $\mathcal{H}$-space), we see that the
associated dipole is a function (three complex components) on the complex
Minkowski space (or $\mathcal{H}$-space). \ The vanishing set of this
function (generically) determines the complex world-line that is referred to
as the complex center of charge or mass. The idea is then to express the
moments in terms of the complex world-line - or as an alternative, find the
complex world-line in terms of the complex dipole. To impliment this (in
principle straightforward) procedure is in practice rather involved,
requiring severe approximations and Clebsch-Gordon expansions of spherical
harmonic products. We illustrate the procedure in detail with the Maxwell
field in flat space and then report the results (obtained earlier) for the
Einstein and Einstein-Maxwell cases with a minimum of detail.

\subsection{Maxwell Fields in Minkowski space}

Beginning with a complex world-line in $\mathbb{M}_{\mathbb{C}}$,$\
z^{a}=\xi ^{a}(\tau ),\ \ $its family of cuts of $\mathfrak{I}_{\mathbb{C}}^{+}\ $is,
as discussed earlier,%
\begin{eqnarray}
u &\equiv &u_{B}=c^{-1}\xi ^{a}(\tau )\widehat{l}_{a}(\zeta ,\overline{\zeta 
})\equiv \frac{\sqrt{2}\tau }{2}-\frac{1}{2}c^{-1}\xi ^{i}Y_{1i}^{0},
\label{u and L} \\
L(u,\zeta ,\overline{\zeta }) &=&c^{-1}\xi ^{a}(\tau )\widehat{m}_{a}(\zeta ,%
\overline{\zeta })=c^{-1}\xi ^{i}(\tau )Y_{1i}^{1}  \notag
\end{eqnarray}%
with $L\ $the angle field of its null normals. Note that $c\ $has been
explicitly$\ $reintroduced so that the cut function $u,\ $with $u_{ret}\ $and%
$\ \tau ,\ $have the dimensions of time$.\ $This has the annoying affect of
causing the frequent appearance of $c.$

\begin{remark}
To avoid a plethora of terms involving $\sqrt{2}\ $we switch from the Bondi
time $u\equiv u_{B},\ $to the retarded time, $u_{ret}=\sqrt{2}u_{B}\ $so that%
\begin{equation}
u_{ret}=\tau -\frac{\sqrt{2}}{2}c^{-1}\xi ^{i}Y_{1i}^{0}  \label{uret}
\end{equation}
Derivatives with respect to $u_{ret}$ are denoted by a prime: $\partial_{u_{ret}}F=F^{\prime}$.
\end{remark}

We now illustrate how an asymptotically flat Maxwell field with
non-vanishing charge determines a unique complex center of charge
world-line, $\xi ^{a}(\tau ).$

We have, first, the asymptotic solution%
\begin{eqnarray}
\phi _{0} &=&\frac{\phi _{0}^{0}}{r^{3}}+O(r^{-4}),  \label{Max1} \\
\phi _{1} &=&\frac{\phi _{1}^{0}}{r^{2}}+O(r^{-3}),  \notag \\
\phi _{2} &=&\frac{\phi _{2}^{0}}{r}+O(r^{-2}).  \notag
\end{eqnarray}%
with the spherical harmonic decomposition%
\begin{eqnarray}
\phi _{0}^{0} &=&\phi _{0i}^{0}Y_{1i}^{1}+\phi _{0ij}^{0}Y_{2ij}^{1}+...,
\label{harmonic decomposition} \\
\phi _{1}^{0} &=&q+\phi _{1i}^{0}Y_{1i}^{0}+\phi _{1ij}^{0}Y_{2ij}^{0}+...,
\\
\phi _{2}^{0} &=&\phi _{2i}^{0}Y_{1i}^{-1}+\phi _{2ij}^{0}Y_{2ij}^{-1}+...,
\end{eqnarray}%
and physical identifications%
\begin{eqnarray}
\phi _{0}^{0} &=&2q\eta ^{i}(u_{ret})Y_{1i}^{1}+c^{-1}Q_{\mathbb{C}%
}^{ij\prime }(u_{ret})Y_{2ij}^{1}+...  \label{Identifications} \\
\phi _{1}^{0} &=&q+\sqrt{2}qc^{-1}\eta ^{i\prime }(u_{ret})Y_{1i}^{0}+\frac{%
\sqrt{2}}{6}c^{-2}Q_{\mathbb{C}}^{ij\prime \prime }(u_{ret})Y_{2ij}^{0}+... 
\notag \\
\phi _{2}^{0} &=&-2qc^{-2}\eta ^{i\prime \prime }(u_{ret})Y_{1i}^{-1}-\frac{1%
}{3}c^{-3}Q_{\mathbb{C}}^{ij\prime \prime \prime }(u_{ret})Y_{2ij}^{-1}+...\
\ \   \notag
\end{eqnarray}%
The quantities $q\eta ^{i}=D_{E\& M}^{i}=D_{E}^{i}+iD_{M}^{i}\ $and $Q_{%
\mathbb{C}}^{ij}\ $are respectively the complex (electric and magnetic)
dipole and complex quadrupole.

Under the null tetrad rotation, Eq.(\ref{null.rotII})%
\begin{eqnarray}
l^{\mathbf{a}} &\rightarrow &l_{C}^{\,\ast a~}=l^{a}-c\frac{\widetilde{L}}{r}%
m^{a}-c\frac{L}{r}\overline{m}^{a}+O(r^{-2}),\   \label{null tetrad rot} \\
m^{\ast a} &=&m^{a}-c\frac{\widetilde{L}}{r}n^{a}, \\
n^{\ast a} &=&n^{a},
\end{eqnarray}%
the leading Maxwell field terms transform as%
\begin{eqnarray}
\phi _{0}^{\ast 0} &=&\phi _{0}^{0}-2cL\ \phi _{1}^{0}+c^{2}L^{2}\phi
_{2}^{0},  \label{null rot} \\
\phi _{1}^{\ast 0} &=&\phi _{1}^{0}-cL\ \phi _{2}^{0},  \notag \\
\phi _{2}^{\ast 0} &=&\phi _{2}^{0},  \notag
\end{eqnarray}

The procedure to determine $\xi ^{a}(\tau )\ $is the following:

In the first equation of Eq.(\ref{null rot}), written as
\begin{equation}
\phi _{0}^{0}=\phi _{0}^{\ast 0}+2cL\phi _{1}^{0}-c^{2}L^{2}\phi _{2}^{0},
\label{basic.I}
\end{equation}%
replace the $u_{ret}$ (appearing in $\phi _{0}^{0},\ \phi _{1}^{0}\ $and $%
\phi _{2}^{0}$) by $u_{ret}=\tau -\frac{\sqrt{2}}{2}c^{-1}\xi
^{i}Y_{1i}^{0}\ $from (\ref{u and L}), and for fixed $\tau ,\ $assume that
the $l=1\ $terms in $\phi _{0}^{\ast 0}\ $vanish.

Formally, by extracting the remaining $l=1\ $terms in Eq.(\ref{basic.I}) via
the integral at constant $\tau ,$%
\begin{equation}
\oint_{S^{2}}\phi _{0}^{0}Y_{1i}^{-1}dS=\oint_{S^{2}}(2cL\phi
_{1}^{0}-c^{2}L^{2}\phi _{2}^{0})Y_{1i}^{-1}dS,  \label{basic.II}
\end{equation}%
we have the exact functional relationship between the dipole $q\eta ^{i}\ $%
and the world-line $\xi ^{a}(\tau )$.

Unfortunately, it is extremely difficult to get explicit relations from Eq.(%
\ref{basic.II}) and approximations applied to Eq.(\ref{basic.I}) must be
used. Our basic approximation is to consider the $\xi ^{a}(\tau )\ $to be of
the form $\xi ^{a}(\tau )=(\tau ,\xi ^{i}(\tau ))\ $(using a $\tau $
re-scaling of the form $\tau \rightarrow F(\tau )$) with both$\ \xi ^{i}\ $%
and the $\eta ^{i}$ to be \textquotedblleft small.\textquotedblright\ We
retain only terms up to second order and harmonic expansions up to $l=2$.

Writing out Eq.(\ref{basic.I}), 
\begin{equation*}
\phi _{0}^{0}=\phi _{0}^{\ast 0}+2cL\phi _{1}^{0}-c^{2}L^{2}\phi _{2}^{0},
\end{equation*}%
using Eqs.(\ref{Identifications}), with $u_{ret}=\tau -\frac{\sqrt{2}}{2}%
c^{-1}\xi ^{i}(\tau )Y_{1i}^{0}\ $and $L=c^{-1}\xi ^{i}(\tau )Y_{1i}^{1},\ $%
by omitting cubic terms including $L^{2}\phi _{2}^{0}\ $, then using the
first two terms of the Taylor series%
\begin{equation}
F\left( {\large \tau }-\frac{\sqrt{2}}{2}c^{-1}{\large \xi }%
^{i}Y_{1i}^{(0)}\right) =F({\large \tau })-\frac{\sqrt{2}}{2}c^{-1}{\large %
\xi }^{i}Y_{1i}^{0}F^{\prime }({\large \tau }),  \label{Taylor}
\end{equation}
and the Clebsch-Gordon expansions of the products of the spherical harmonics
(see Appendix \ref{A}) we finally have (after simplification) for just the $l=1$
harmonic terms 
\begin{equation}
q\eta ^{k}=q\xi ^{k}-i\frac{q}{2}\xi ^{l}c^{-1}\eta ^{i\prime }\epsilon
_{ilk}+\frac{\sqrt{2}}{10}c^{-2}Q_{\mathbb{C}}^{ik\prime \prime }\xi ^{i}.
\label{result.I}
\end{equation}

The first thing we notice is the linear relation:%
\begin{equation}
\eta ^{j}(\tau )={\large \xi }^{j}(\tau ).  \label{linear}
\end{equation}

This can be fed back into Eq.(\ref{result.I}) in either of two ways
resulting in either of the relations:

\begin{eqnarray}
\eta ^{k} &=&\xi ^{k}-i\frac{1}{2}c^{-1}\xi ^{l}\xi ^{i\prime }\epsilon
_{ilk}+\frac{\sqrt{2}}{10}q^{-1}c^{-2}Q_{\mathbb{C}}^{ik\prime \prime }\xi
^{i},  \label{result.II} \\
\xi ^{k} &=&\eta ^{k}+i\frac{1}{2}c^{-1}\eta ^{l}\eta ^{i\prime }\epsilon
_{ilk}-\frac{\sqrt{2}}{10}q^{-1}c^{-2}Q_{\mathbb{C}}^{ik\prime \prime }\xi
^{i},  \notag
\end{eqnarray}%
which determine the complex dipole in terms of the complex world-line or the
world-line in terms of the complex dipole.

Note that though all the expressions are functions of $\tau ,$\ the $\tau \ $%
can be replaced by $u_{ret}\ $with no other changes needed due to our
approximation scheme.

\subsection{Asymptotically Flat Space-Times}

Turning now to the Einstein (or Einstein-Maxwell) case, we basically repeat
the procedure used in the Minkowski space Maxwell field example.

We begin with an unknown complex world-line in $\mathcal{H}$-space,$\
z^{a}=\xi ^{a}(\tau ),\ $to be determined by the existing asymptotically
flat space-time. $\ $Its family of cuts and null normal angle field of $%
\mathfrak{I}_{\mathbb{C}}^{+}\ $is, as discussed earlier (c.f., (\ref%
{solution*}), (\ref{GCEq}), etc.), given by%
\begin{eqnarray}
u &=&c^{-1}G(\tau ,\zeta ,\overline{\zeta })  \label{cut**} \\
&=&c^{-1}\xi ^{a}(\tau )\hat{l}_{a}(\zeta ,\overline{\zeta }%
)+c^{-1}H_{l\geqslant 2}(\xi ^{a}(\tau ),\zeta ,\overline{\zeta })=c^{-1}\xi
^{a}(\tau )\hat{l}_{a}(\zeta ,\overline{\zeta })+c^{-1}\tilde{H}_{l\geqslant
2}(\tau ,\zeta ,\overline{\zeta }),  \notag \\
&=&\frac{\sqrt{2}\tau }{2}-\frac{1}{2}c^{-1}\xi ^{i}(\tau
)Y_{1i}^{0}+c^{-1}\xi ^{ij}(\tau )Y_{2ij}^{0}+\ldots   \notag \\
L(u,\zeta ,\overline{\zeta }) &=&c^{-1}\eth _{\tau }G(\tau ,\zeta ,\overline{%
\zeta })|_{\tau =T(u,\zeta ,\overline{\zeta })}=c^{-1}\xi ^{i}(\tau
)Y_{1i}^{1}-6c^{-1}\xi ^{ij}(\tau )Y_{1ij}^{1}+\ldots   \label{L**} \\
\sigma ^{0}(\tau ,\zeta ,\bar{\zeta}) &=&\eth _{(\tau )}^{2}G(\tau ,\zeta ,%
\bar{\zeta})=24\xi ^{ij}(\tau )Y_{2ij}^{2}+\ldots   \label{sigma**}
\end{eqnarray}

We now show how a given asymptotically flat space-time determines the
complex center of charge world-line $\xi ^{a}(\tau ).$

Returning to the ``peeling'' theorem:

\begin{eqnarray*}
\psi _{0} &=&\psi _{0}^{0}r^{-5}+O(r^{-6}), \\
\psi _{1} &=&\psi _{1}^{0}r^{-4}+O(r^{-5}), \\
\psi _{2} &=&\psi _{2}^{0}r^{-3}+O(r^{-4}), \\
\psi _{3} &=&\psi _{3}^{0}r^{-2}+O(r^{-3}), \\
\psi _{4} &=&\psi _{4}^{0}r^{-1}+O(r^{-2}),
\end{eqnarray*}%
with the transformation law of the leading terms under a null rotation,

\begin{eqnarray}
\psi _{0}^{\ast 0} &=&\psi _{0}^{0}-4cL\psi _{1}^{\ast 0}+6c^{2}L^{2}\psi
_{2}^{\ast 0}-4c^{3}L^{3}\psi _{3}^{\ast 0}+c^{4}L^{4}\psi _{4}^{0},
\label{a} \\
\psi _{1}^{\ast 0} &=&\psi _{1}^{0}-3cL\psi _{2}^{0}+3c^{2}L^{2}\psi
_{3}^{0}-c^{3}L^{3}\psi _{4}^{0},  \label{b} \\
\psi _{2}^{\ast 0} &=&\psi _{2}^{0}-2cL\psi _{3}^{0}+c^{2}L^{2}\psi _{4}^{0},
\label{c} \\
\psi _{3}^{\ast 0} &=&\psi _{3}^{0}-cL\psi _{4}^{0},  \label{d} \\
\psi _{4}^{\ast 0} &=&\psi _{4}^{0},  \label{e}
\end{eqnarray}%
we can then determine the transformation law for the physical quantities
that are identified in the following harmonic components.

The (truncated) harmonic expansions with their (approximate) physical
identifications are%
\begin{eqnarray}
\psi _{0}^{0} &=&\psi _{0}^{0ij}Y_{2ij}^{2}+...  \label{harmonic expansion}
\\
\psi _{1}^{0} &=&\psi _{1}^{0i}Y_{1i}^{1}+...  \notag \\
\psi _{2}^{0} &=&\Psi -\eth ^{2}\bar{\sigma}^{0}-c^{-1}\sigma ^{0}(\bar{%
\sigma}^{0})^{\prime }  \notag \\
\Psi  &=&\overline{\Psi }=\Psi ^{0}+\Psi ^{i}Y_{1i}^{0}+...  \notag \\
\psi _{3}^{0} &=&c^{-1}\eth \overline{\sigma }^{0 \prime }  \notag \\
\psi _{4}^{0} &=&-c^{-2}\overline{\sigma }^{0 \prime \prime }  \notag
\end{eqnarray}%
and 
\begin{eqnarray}
\psi _{0}^{0ij} &=&\text{approximately, the quadrupole}
\label{identifications} \\
D_{\mathbb{C(}grav\mathbb{)}}^{i} &=&D_{(mass)}^{i}+ic^{-1}J^{i}=-\frac{c^{2}%
\sqrt{2}}{12G}\psi _{1}^{0i}  \label{ID2} \\
\psi _{1}^{0i} &=&-\frac{6\sqrt{2}G}{c^{2}}(D_{(mass)}^{i}+ic^{-1}J^{i})
\label{ID3} \\
\Psi  &\equiv &\psi _{2}^{0}+\eth ^{2}\overline{\sigma }^{0}+c^{-1}\sigma
^{0}\dot{\overline{\sigma }}^{0} \\
\Psi  &=&\overline{\Psi }=\Psi ^{0}+\Psi ^{i}Y_{1i}^{0}+...\ \ \ \ \ \ \ \ 
\label{ID4} \\
&=&-M_{B}\frac{2\sqrt{2}G}{c^{2}}-\frac{6G}{c^{3}}P^{i}Y_{1i}^{0}+...\ \ 
\text{Mass Aspect}  \notag \\
M_{B} &=&-\frac{c^{2}}{2\sqrt{2}G}\Psi ^{0},\ \text{Bondi mass}  \label{ID5}
\\
P^{i} &=&-\frac{c^{3}}{6G}\Psi ^{i},\ \ \ \text{Bondi linear momentum}
\label{ID6} \\
\sigma ^{0} &=&24\xi ^{ij}Y_{2ij}^{2}+\ldots \ \ \text{Bondi asymptotic
shear.}  \label{ID7} \\
\xi ^{ij} &=&(\xi _{R}^{ij}+i\xi _{I}^{ij})=\frac{G}{12\sqrt{2}c^{4}}%
(Q_{\mathrm{Mass}}^{ij\prime \prime }+iQ_{\mathrm{Spin}}^{ij\prime \prime })  \label{ID8}
\end{eqnarray}

\begin{remark}
The relationship between $\psi _{1}^{0i}\ $and the mass dipole, $%
D_{(mass)}^{i},$ and angular momentum, $J^{i},$ is usually considered to be
more complicated than stated here, often involving quadratic terms in the
Bondi shear \cite{Szabados:2004}. The trouble is that there are disagreements in these quadratic
terms in the different versions. We have simply left them out here and note
that in our approximations the disputed terms do not appear. \ 
\end{remark}

Concentrating on the transformation of the dipole, Eqs.(\ref{a})-(\ref{e}),
our focus is only on Eq.(\ref{b}), which can be rewritten as%
\begin{equation}
\psi _{1}^{0}=\psi _{1}^{\ast 0}+3cL\psi _{2}^{0}-3c^{2}L^{2}\psi
_{3}^{0}+c^{3}L^{3}\psi _{4}^{0}.  \label{basic*}
\end{equation}%
We first note that the $l=1\ $term in $\psi _{1}^{0}\ $is proportional to
the complex gravitational dipole, $D_{\mathbb{C}(grav)}^{i}(u).\ $%
Our procedure is now to replace all the $u$s that appear in Eq.(\ref%
{basic*}) by$\ u_{ret}=(\sqrt{2}c)^{-1}\xi ^{a}(\tau )\hat{l}_{a}(\zeta ,\overline{\zeta }%
)+(\sqrt{2})^{-1}H_{l\geqslant 2}(\xi ^{a}(\tau ),\zeta ,\overline{\zeta })$ (i.e., Eq.(\ref%
{cut**})), and remember that $L\ $also$\ $depends on $\xi ^{a}(\tau )$.
Then, from our basic assumption, we take the $l=1\ $term in $\psi _{1}^{\ast
0}\ $to vanish and finally extract the $l=1\ $harmonic coefficients from Eq.(%
\ref{cut**}). Formally, this is done via the integral expression%
\begin{equation}
\oint_{S^{2}}\psi _{1}^{0}Y_{1i}^{-1}dS=\oint_{S^{2}}(3cL\psi
_{2}^{0}-3c^{2}L^{2}\psi _{3}^{0}+c^{3}L^{3}\psi _{4}^{0})Y_{1i}^{-1}dS,
\label{basic**}
\end{equation}%
which on the left-side contains $\tau \ $and the dipole $D_{\mathbb{C}%
(grav)}^{i}\ $while the right-side contains $\tau ,\ $the unknown world-line 
$\xi ^{a}(\tau )\ $and the Bondi shear, $\sigma^{0}$.

Though in principle this equation should allow us to establish the
relationship between $D_{\mathbb{C}(grav)}^{i}\ \ $and $\xi ^{a}(\tau ),\ $%
in practice this is not possible: we must to return to Eq.(\ref{basic*}) and
use harmonic and Clebsch-Gordon expansions with severe approximations and
finally collect the $l=1\ $terms directly.

\subsubsection{A Poor Approximation: Results}

We first describe a preliminary procedure for extracting the complex
world-line from an asymptotically flat space-time rather explicitly. The
following approximations are used: the Bondi mass is taken as zeroth order
while all other variables are first-order with the calculations done keeping
terms up to second order. In this preliminary version, the harmonic
expansions keep only the $l=(0,1)\ $terms. This implies the severe condition
that the Bondi shear ($\sigma ^{0}$) be taken to be zero. Later this
condition is relaxed.

Via these approximations Eq.(\ref{basic*}) becomes

\begin{equation}
\psi _{1}^{0}(u,\zeta ,\overline{\zeta })=\psi _{1}^{\ast 0}+3cL(\tau ,\zeta
,\overline{\zeta })\Psi (u,\zeta ,\overline{\zeta }).  \label{basic***}
\end{equation}%
\qquad

Using the retarded time $u_{ret}=\sqrt{2}u\ $instead of the Bondi time, then
replacing all the $u_{ret}$s by $u_{ret}=\tau -\frac{\sqrt{2}}{2}c^{-1}\xi
^{i}(\tau )Y_{1i}^{0}\ $and finally Taylor expanding with (\ref{Taylor}), $\ 
$Eq. (\ref{basic***}) becomes (with physical identifications inserted):

\begin{equation}
\psi _{1}^{0i}(\tau )Y_{1i}^{1}-\frac{\sqrt{2}}{2}c^{-1}\psi _{1}^{0i\prime
}(\tau )\xi ^{j}Y_{1j}^{0}Y_{1i}^{1}=-\frac{6\sqrt{2}G}{c^{2}}M_{B}\xi
^{i}(\tau )Y_{1i}^{1}-\frac{18G}{c^{3}}P^{i}\xi ^{j}(\tau
)Y_{1i}^{0}Y_{1j}^{1}.  \label{basic****}
\end{equation}

Finally after Clebsch-Gordon expansions and the use of the (linearized)
Bianchi identity, Eq.(\ref{AsyBI1*}), 
\begin{equation}
\sqrt{2}\psi _{1}^{0\prime }=2c\Psi ^{i}Y_{1i}^{1}\ \ \ \ \Rightarrow \ \
\psi _{1}^{0i\,\prime }=\sqrt{2}c\Psi ^{i}=-\frac{6\sqrt{2}G}{c^{2}}P^{i},
\label{BI}
\end{equation}%
the three complex $l=1\ $coefficients of (\ref{basic****}), (with $\xi
^{k}=\xi _{R}^{k}+i\xi _{I}^{k}$), yield 
\begin{equation}
(D_{(mass)}^{k}+ic^{-1}J^{k})=M_{B}(\xi _{R}^{k}+i\xi _{I}^{k})-\epsilon
_{ijk}c^{-1}P^{i}(i\xi _{R}^{j}-\xi _{I}^{j}),  \label{result 0}
\end{equation}%
or the pair of real equations,%
\begin{eqnarray}
D_{(mass)}^{k}(\tau ) &=&M_{B}\xi _{R}^{k}(\tau )+c^{-1}\epsilon
_{ijk}P^{i}\xi _{I}^{j},  \label{result 1} \\
J^{k}(\tau ) &=&cM_{B}\xi _{I}^{k}(\tau )+\epsilon _{ijk}P^{j}\xi
_{R}^{i}(\tau ).  \label{result 2}
\end{eqnarray}

%\begin{remark}
%In an earlier paper \{REf.LR\}, using a very much more complicated
%derivation procedure, the same complex mass dipole was derived.
%Unfortunately an error in Eq.(302), 
%\begin{equation*}
%D_{(\mathrm{mass})}^{\,j}=M_{\mathrm{B}}\xi _{R}^{j}+c^{-1}(M_{\mathrm{B}%
%}\xi _{R}^{k\,\prime }\xi _{I}^{i}+\frac{1}{2}M_{\mathrm{B}}\xi _{R}^{k}\xi
%_{I}^{i\,\prime })\epsilon _{kij},
%\end{equation*}%
%appears to have been made in that work. The last term, $\frac{1}{2}M_{%
%\mathrm{B}}\xi _{R}^{k}\xi _{I}^{i\,\prime }\epsilon _{kij},$ is almost
%certainly wrong. \qquad
%\end{remark}

From Eq.(\ref{BI}), we immediately get the kinematic definition of the Bondi
momentum in terms of the complex world-line. In addition, we have the
conservation of angular-momentum which arises from the reality$\ $of the
mass aspect, $\Psi ,\ $Eq.(\ref{ID4})$:$%
\begin{eqnarray}
P^{i} &=&D_{(mass)}^{i\prime }=M_{B}\xi _{R}^{k\ \prime }+c^{-1}\epsilon
_{ijk}M_{B}(\xi _{R}^{i\ \prime }\xi _{I}^{j})^{\prime },  \label{momentum}
\\
J^{k\ \prime } &=&0.  \label{ang.Mom.cons}
\end{eqnarray}

There are several things of significance that should be pointed out here.

\begin{itemize}
\item If the higher gravitational moments and electromagnetic terms were
included, these results, Eqs.(\ref{result 1})-(\ref{ang.Mom.cons}), would
all be augmented by further terms. \ In particular there would be a
non-vanishing angular-momentum flux. See below.

\item In the expression for the angular momentum there are two terms, the
second being the conventional orbital angular momentum while the first has
been identified, via the Kerr metric and the charged Kerr metric
\cite{KerrNewman65}, as the intrinsic spin angular momentum.

\item The mass dipole contains the conventional $M\overrightarrow{R}\ \ $plus a
momentum-spin interaction term that creates a spin-velocity coupling contribution to the linear momentum.
\end{itemize}

These results - basically kinematic, aside from the conservation of
angular-momentum - have been derived, with severe approximations, by
associating the idea of a complex center of mass curve with a complex curve
in $\mathcal{H}$-space.

In the same vein (with the same approximations), we obtain the dynamic law
for the motion of the real part of $\xi ^{k\ }$ (i.e., $\xi _{R}^{k\ }$).$\ $%
From the second Bianchi identity, Eq.(\ref{AsyBI2*})%
\begin{equation}
\sqrt{2}\psi _{2}^{0\prime }=-c\eth \psi _{3}^{0}+c\sigma ^{0}\psi _{4}^{0}\
\ \Rightarrow \ \psi _{2}^{0\prime }=0\   \label{BI2}
\end{equation}%
we have that $M_{B}\ $and $P^{i}$\ are constant, i.e. conservation of energy
and momentum:

\begin{equation}
M_{B}^{\prime }=P^{i\prime }=0.  \label{conservation}
\end{equation}

\subsubsection{A Better Approximation (with Bondi shear): Results}

For a more accurate description/determination of the complex world-line
associated with a given asymptotically flat Einstein (or in the following subsection, Einstein-Maxwell)
space-time we restore, in the calculations, the Bondi shear and include the
effects of the Einstein-Maxwell equations. Rather than redoing the
calculations from the beginning, using the same procedures as in the
previous section, we simply give the final results. The approximations are
basically the same: the Bondi mass is zero-order, while all other variables
are first -order; in the calculations only quadratic terms are retained.
However the harmonic expansions now include the $l=(0,1,2)$ harmonics.
In addition, since a Maxwell field (with non-vanishing total charge) is
allowed, we have, not only the complex center of mass line, $z^{a}=\xi
^{a}(\tau ),\ $but at well the complex center of charge line. It is denoted
by $z^{a}=\eta ^{a}(\tau ).\ $In general the two line are different, though
in special circumstance they can coincide.

The idea is to start with Eq.(\ref{basic***}), use the known expression for $%
\psi _{1}^{0}(u_{ret},\zeta ,\overline{\zeta })\ $and $\Psi (u_{ret},\zeta ,%
\overline{\zeta }),\ $in terms of the physical gravitational moments, then replace every $u_{ret}\ $by
\begin{eqnarray}
u_{ret} &=&\frac{\sqrt{2}}{2}c^{-1}X(\xi ^{a}(\tau ),\zeta ,\overline{\zeta }%
),  \label{ret} \\
&=&\tau -\frac{\sqrt{2}}{2}c^{-1}\xi ^{i}(\tau )Y_{1i}^{0}+\sqrt{2}c^{-1}\xi
^{ij}(\tau )Y_{1ij}^{0}+...  \notag
\end{eqnarray}%
set$\ $the $l=1\ $coefficients of $\psi _{1}^{\ast 0}\ $to zero and finally
extract the $l=1\ \ $coefficients from the entire equation - a long process
involving repeated Clebsch-Gordon expansions (c.f., \cite{Kozameh:2008gw, Adamo:2009vu}). This process leads to

\begin{eqnarray}
\psi _{1}^{0i} &=&-6\sqrt{2}Gc^{-2}(D_{(mass)}^{i}+ic^{-1}J^{i})
\label{complex grav.dipole} \\
(D_{(mass)}^{i}+ic^{-1}J^{i}) &=&M_{B}\xi ^{i}+ic^{-1}P^{j}\xi ^{l}\epsilon
_{lji}-\frac{4G}{5c^{5}}P^{i}Q_{\mathrm{Grav}}^{ij\,\prime \prime }-\frac{3%
}{5c^{2}}\overline{Q}_{\mathrm{Grav}}^{il\prime \prime }\xi ^{l} \notag \\ & & -i\frac{3G}{5c^{6}}Q_{\mathrm{Grav}}^{lk\prime \prime }\overline{Q}_{\mathrm{Grav}%
}^{kj\prime \prime }\epsilon _{lji}  \notag
\end{eqnarray}
or from the real and imaginary parts,
\begin{eqnarray}
D_{(mass)}^{i} &=&M_{B}\xi _{R}^{i}+c^{-1}\xi _{I}^{l}P^{j}\epsilon _{ijl}-%
\frac{4G}{5c^{5}}P^{i}Q_{\mathrm{Mass}}^{ij\,\prime \prime } \notag \\
& & -\frac{3}{5c^{2}}(Q_{\mathrm{Mass}}^{il\prime \prime }\xi _{R}^{l}+Q_{\mathrm{Spin}%
}^{il\prime \prime }\xi _{I}^{l})-\frac{6G}{5c^{6}}Q_{\mathrm{Mass}%
}^{kl\prime \prime }Q_{\mathrm{Spin}}^{kj\prime \prime }\epsilon _{lji} 
\notag \\
J^{i} &=&M_{B}c\xi _{I}^{i}+P^{j}\xi _{R}^{l}\epsilon _{lji}-\frac{4G}{5c^{5}}P^{i}Q_{\mathrm{Spin}}^{ij\,\prime \prime }-\frac{3}{5c}(Q_{%
\mathrm{Mass}}^{il\prime \prime }\xi _{I}^{l}-Q_{\mathrm{Spin}}^{il\prime
\prime }\xi _{R}^{l}).  \notag
\end{eqnarray}%
the definition of the mass dipole and angular momentum in terms of the
complex world-line

The kinematic definition of the linear momentum and the angular momentum
conservation law are then found by extracting the $l=1\ $harmonics from the
Bianchi identity, Eq.(\ref{AsyBI1*}), (the evolution equation for $\psi
_{1}^{0}$) 
\begin{equation*}
\dot{\psi}_{1}^{0}=-\eth \psi _{2}^{0}+2\sigma ^{0}\psi _{3}^{0},
\end{equation*}
or
\begin{equation}\label{AsyBI1}
\psi _{1}^{0\,\prime } =-\frac{\sqrt{2}}{2}c\eth \Psi +\frac{\sqrt{2}}{2}c%
\eth ^{3}\overline{\sigma }+3\sigma ^{0}\eth (\overline{\sigma }^{\prime }).
\end{equation}%
leading to
\begin{equation*}
(D_{(mass)}^{i\prime }+ic^{-1}J^{i\prime })=P^{i}+i\frac{12G}{5c^{6}}Q_{%
\mathrm{Grav}}^{kl\,\prime \prime }\overline{Q}_{\mathrm{Grav}}^{lj\,\prime
\prime \prime }\epsilon _{jki}.
\end{equation*}

Then inserting the expressions for $D_{(mass)}^{i}$and $J^{i}$ we obtain,
from the real part, an expression for the linear momentum

\begin{eqnarray}
P^{i} &=&D_{(mass)}^{i\ \prime }-\frac{12G}{5c^{6}}\left(Q_{\mathrm{Mass}%
}^{kl\,\prime \prime }Q_{\mathrm{Spin}}^{lj\,\prime \prime }\right)^{\prime
}\epsilon _{jki},  \label{P} \\
P^{i} &=&M_{B}\xi _{R}^{i\prime }+\mathfrak{P}^{i},  \label{P*} \\
\mathfrak{P}^{i} &=&c^{-1}(\xi _{I}^{l}P^{j})^{\prime }\epsilon _{jli}-\frac{%
4G}{5c^{5}}(P^{i}Q_{\mathrm{Mass}}^{ij\,\prime \prime })^{\prime}-\frac{3}{5c^{2}}(Q_{\mathrm{Mass}}^{il\prime \prime }\xi _{R}^{l}+Q_{\mathrm{Spin}}^{il\prime \prime
}\xi _{I}^{l})^{\prime } \notag \\ 
& & -\frac{3G}{c^{6}}\left( Q_{\mathrm{Mass}}^{kl\,\prime \prime
}Q_{\mathrm{Spin}}^{lj\,\prime \prime }\right) ^{\prime }\epsilon _{jki}, 
\notag
\end{eqnarray}%
and, from the imaginary part, the angular momentum conservation law:

\begin{eqnarray}
&&J^{i\,\prime }=(\mathrm{Flux})^{i},  \label{ang.mom.flux} \\
J^{i} &=&M_{B}c\xi _{I}^{i}+P^{j}\xi _{R}^{l}\epsilon _{lji}-\frac{4G}{5c^{5}}
P^{i}Q_{\mathrm{Spin}}^{ij\,\prime \prime }-\frac{3}{5c}%
(Q_{Mass}^{il\prime \prime }\xi _{I}^{l}-Q_{Spin}^{il\prime \prime
}\xi _{R}^{l}),  \label{J^T} \\
(\mathrm{Flux})^{i} &=&\frac{12G}{5c^{5}}\left( Q_{\mathrm{Mass}}^{lk\,\prime \prime
}Q_{\mathrm{Mass}}^{jl\,\prime \prime \prime }+Q_{\mathrm{Spin}}^{kl\,\prime
\prime }Q_{\mathrm{Spin}}^{jl\,\prime \prime \prime }\right) \epsilon _{ijk}.
\label{flux*}
\end{eqnarray}

Finally, from the $l=0,1$ parts of the Bianchi identity, Eq.(\ref{AsyBI2*}),
the evolution equation for the mass aspect 
\begin{equation*}
\dot{\psi}_{2}^{0} =-\eth \psi _{3}^{0}+\sigma ^{0}\psi _{4}^{0},
\end{equation*}
or
\begin{equation}\label{evolution} \\
\Psi ^{\prime } =\frac{\sqrt{2}}{c}\sigma ^{0\prime }\overline{\sigma }%
^{0\prime },
\end{equation}%
we obtain both the energy loss expression and the evolution of the momentum,
i.e., the equations of motion.

The energy (mass) loss equation, from the $l=0$ part, is 
\begin{equation*}
M_{B}^{\prime }=-\frac{G}{5c^{7}}\left( Q_{\mathrm{Mass}}^{ij\,\prime \prime
}Q_{\mathrm{Mass}}^{ij\,\prime \prime \prime }+Q_{\mathrm{Spin}}^{ij\,\prime
\prime }Q_{\mathrm{Spin}}^{ij\,\prime \prime \prime }\right) ,
\end{equation*}%
the known quadrupole expression, while the momentum loss equation, from the $%
l=1$ part of (\ref{evolution}) becomes a version of Newton's second law: 
\begin{eqnarray}
P^{k\,\prime } &=&F_{\mathrm{recoil}}^{k},  \label{recoil} \\
F_{\mathrm{recoil}}^{k} &\equiv &\frac{2G}{15c^{6}}{\large (}Q_{\mathrm{Spin}%
}^{lj\,\prime \prime \prime }Q_{\mathrm{Mass}}^{ij\,\prime \prime \prime
}-Q_{\mathrm{Mass}}^{lj\,\prime \prime \prime }Q_{\mathrm{Spin}}^{ij\,\prime
\prime \prime }{\large )}\epsilon _{ilk}.  \notag
\end{eqnarray}

Finally substituting the $P^{i}$ from Eq.(\ref{P}), we have Newton's 2nd law
of motion;

\begin{equation}
M_{B}\xi _{R}^{i\,\prime \prime }=F_{\mathrm{recoil}}^{k}-M_{B}^{\ \prime
}\xi _{R}^{i\prime }-\mathfrak{P}^{i\prime }\equiv F^{i}  \label{newton2nd}
\end{equation}

\subsubsection{Results for Einstein-Maxwell Space-times}

The calculations that were performed earlier for the vacuum GR case can be
extended to the Einstein-Maxwell case with considerably more effort. Rather
than going into the details we will simply present the main results. The
Maxwell field considered has only charge and dipole terms: with a bit of
effort quadrupole terms could be included. The main change needed
is the modification of the asymptotic Bianchi identities to include the
Maxwell field:%
\begin{eqnarray}
\dot{\psi}_{2}^{0} &=&-\eth \psi _{3}^{0}+\sigma ^{0}\psi _{4}^{0}+k\phi
_{2}^{0}\bar{\phi}_{2}^{0},  \label{ABI*1} \\
\dot{\psi}_{1}^{0} &=&-\eth \psi _{2}^{0}+2\sigma ^{0}\psi _{3}^{0}+2k\phi
_{1}^{0}\bar{\phi}_{2}^{0},  \label{AsyBI2} \\
k &=&2Gc^{-4}.  \label{k}
\end{eqnarray}%
where the fields $\phi _{1}^{0}\ $and $\phi _{2}^{0}\ $are given by Eqs.(\ref%
{Identifications}).

The result of the calculations are 
\begin{eqnarray}
\psi _{1}^{0i} &=&-\frac{6\sqrt{2}G}{c^{2}}(D_{(mass)}^{i}+ic^{-1}J^{i}),
\label{complex dipole*} \\
(D_{(mass)}^{k}+ic^{-1}J^{k}) &=&M_{B}\xi ^{k}+\frac{i\epsilon _{mik}}{c}\xi
^{m}P^{i}-i\frac{q^{2}}{3c^{2}}\epsilon _{mik}\xi ^{m}\overline{\eta }%
^{i\,\prime \prime }-\frac{4G}{5c^{5}}P^{i}Q_{\mathrm{Grav}}^{ik\prime
\prime }  \label{C.D*} \\
&&+\frac{\sqrt{2}Gq^{2}}{15c^{6}}\overline{\eta }^{j\,\prime \prime }Q_{%
\mathrm{Grav}}^{kj}-\frac{3}{5c^{2}}\xi ^{j}\overline{Q}_{\mathrm{Grav}%
}^{kj\prime \prime }-i\frac{3G}{5c^{6}}\epsilon _{mjk}Q_{\mathrm{Grav}%
}^{im\prime \prime }\overline{Q}_{\mathrm{Grav}}^{ij\prime \prime }  \notag
\end{eqnarray}%
or
\begin{eqnarray}
D_{(mass)}^{i} &=&M_{B}\xi _{R}^{i}-c^{-1}P^{j}\xi _{I}^{k}\varepsilon
_{kji}-\frac{3}{5c^{2}}\left( \xi _{R}^{j}Q_{\mathrm{Mass}}^{ij\prime \prime
}+\xi _{I}^{j}Q_{\mathrm{Spin}}^{ij\prime \prime }\right)  \label{dipole*} \\  & & +\frac{q^{2}}{3c^{2}}\left( \xi _{I}^{j}\eta _{R}^{k\prime \prime }-\xi _{R}^{j}\eta
_{I}^{k\prime \prime }\right) \epsilon _{jki}-\frac{4G}{5c^{5}}P^{j}Q_{\mathrm{Mass}}^{ij\prime \prime }-\frac{3G}{%
5c^{6}}Q_{\mathrm{Mass}}^{kl\prime \prime }Q_{\mathrm{Spin}}^{kj\prime
\prime }\epsilon _{lji} \notag \\ 
& & +\frac{\sqrt{2}Gq^{2}}{15c^{6}}\left( \eta
_{R}^{j\prime \prime }Q_{\mathrm{Mass}}^{ij\prime \prime }+\eta
_{I}^{j\prime \prime }Q_{\mathrm{Spin}}^{ij\prime \prime }\right)   \notag
\end{eqnarray}%
and%
\begin{eqnarray}
J^{i} &=& cM_{B}\xi _{I}^{i}+\xi _{R}^{k}P^{j}\epsilon _{kji}-\frac{3}{5c}%
\left( Q_{\mathrm{Mass}}^{ij\prime \prime }\xi _{I}^{j}-Q_{\mathrm{Spin}%
}^{ij\prime \prime }\xi _{R}^{j}\right) \label{J*} \\
& & +\frac{q^{2}}{3c}\left( \xi_{R}^{k}\eta _{R}^{j\prime \prime }+\xi _{I}^{k}\eta _{I}^{j\prime \prime}\right) \epsilon _{kji}-\frac{4G}{5c^{4}}P^{j}Q_{\mathrm{Spin}}^{ij\prime \prime }+\frac{\sqrt{2}%
Gq^{2}}{15c^{5}}\left( \eta _{R}^{j\prime \prime }Q_{\mathrm{Spin}%
}^{ij\prime \prime }-\eta _{I}^{j\prime \prime }Q_{\mathrm{Mass}}^{ij\prime
\prime }\right) .  \notag
\end{eqnarray}%
\qquad 

From the Bianchi identity, Eq.(\ref{AsyBI2}), we have%
\begin{equation*}
D_{(mass)}^{i\ \prime }+ic^{-1}J^{i\ \prime }=P^{i}+\frac{2q^{2}}{3c^{3}}%
\bar{\eta}^{i\prime \prime }+\frac{12Gi}{5c^{6}}Q_{\mathrm{Grav}}^{kl\prime
\prime }\bar{Q}_{\mathrm{Grav}}^{lj\prime \prime \prime }\epsilon _{jki}+%
\frac{2\sqrt{2}iq^{2}}{3c^{4}}\eta ^{k\prime }\bar{\eta}^{j\prime \prime
}\epsilon _{jki}
\end{equation*}%
or, from the real part,

\begin{eqnarray}
P^{i} &=&M_{B}\xi _{R}^{i\prime }-\frac{2q^{2}}{3c^{3}}\eta _{R}^{i\prime
\prime }-c^{-1}(P^{j}\xi _{I}^{k})^{\prime }\epsilon _{kji}+\frac{q^{2}}{%
3c^{2}}\left( \xi _{I}^{j}\eta _{R}^{k\prime \prime }-\xi _{R}^{j}\eta
_{I}^{k\prime \prime }\right) ^{\prime }\epsilon _{jki} \label{momEM} \\
& & +\frac{2\sqrt{2}%
q^{2}}{3c^{4}}\left( \eta _{I}^{k\prime }\eta _{R}^{j\prime }\right)
^{\prime }\epsilon _{jki}-\frac{4G}{5c^{5}}\left( P^{j}Q_{\mathrm{Mass}}^{ij\prime \prime }\right)
^{\prime }+\frac{\sqrt{2}Gq^{2}}{15c^{6}}\left( \eta _{R}^{j\prime \prime
}Q_{\mathrm{Mass}}^{ij\prime \prime }+\eta _{I}^{j\prime \prime }Q_{\mathrm{%
Spin}}^{ij\prime \prime }\right) ^{\prime } \notag \\
& & -\frac{3}{5c^{2}}\left( \xi
_{R}^{j}Q_{\mathrm{Mass}}^{ij\prime \prime }+\xi _{I}^{j}Q_{\mathrm{Spin}%
}^{ij\prime \prime }\right) ^{\prime }+\frac{3G}{c^{6}}\left( Q_{\mathrm{Spin}}^{lk\prime \prime }Q_{\mathrm{Mass%
}}^{lj\prime \prime }\right) ^{\prime }\epsilon _{jki}, \notag
\end{eqnarray}%
and imaginary part
\begin{eqnarray}
J^{i\prime} & = & (\mathrm{Flux})^{i} \label{Flux1} \\
(\mathrm{Flux})^{i} & = & \frac{2\sqrt{2}q^{2}}{3c^{3}}\left(\eta^{k\prime}_{R}\eta^{j\prime\prime}_{R}+\eta^{k\prime}_{I}\eta^{j\prime\prime}_{I}
\right)\epsilon_{jki}+\frac{12G}{5c^{5}}\left( Q_{\mathrm{Mass}%
}^{lk\prime \prime }Q_{\mathrm{Mass}}^{lj\prime \prime \prime }+Q_{\mathrm{%
Spin}}^{lk\prime \prime }Q_{\mathrm{Spin}}^{lj\prime \prime \prime }\right)\epsilon_{kji}-\frac{2q^{2}}{3c^{2}}\eta^{i\prime\prime}_{I}. \notag
\end{eqnarray}

Alternatively, we can consider the term $\frac{2q^{2}}{3c^{2}}\eta^{i\prime\prime}_{I}$ as a contribution to the total angular momentum rather than the flux.  Using this alternative definition of angular momentum,
\begin{equation*}
J_{T}^{i}=J^{i}+\frac{2q^{2}}{3c^{2}}\eta _{I}^{i\prime },
\end{equation*}%
where $J^{i}$ is given by Eq.(\ref{J*}), leads to a modified flux law:
\begin{eqnarray}
J_{T}^{i\prime} & = & (\mathrm{Flux})^{i}_{T} \label{FluxT} \\
(\mathrm{Flux})_{T}^{i} & = & \frac{2\sqrt{2}q^{2}}{3c^{3}}\left( \eta _{R}^{k\prime
}\eta _{R}^{j\prime \prime }+\eta _{I}^{k\prime }\eta _{I}^{j\prime \prime
}\right) \epsilon _{jki}+\frac{12G}{5c^{5}}\left( Q_{\mathrm{Mass}%
}^{lk\prime \prime }Q_{\mathrm{Mass}}^{lj\prime \prime \prime }+Q_{\mathrm{%
Spin}}^{lk\prime \prime }Q_{\mathrm{Spin}}^{lj\prime \prime \prime }\right)
\epsilon _{kji}. \notag
\end{eqnarray}%

Finally from the Bianchi identity, Eq.(\ref{ABI*1}), we have the mass loss
and momentum loss equations:%
\begin{equation}\label{ML}
M_{B}^{\prime }=-\frac{G}{5c^{7}}\left( Q_{\mathrm{Mass}}^{ij\prime \prime
\prime }Q_{\mathrm{Mass}}^{ij\prime \prime \prime }+Q_{\mathrm{Spin}%
}^{ij\prime \prime \prime }Q_{\mathrm{Spin}}^{ij\prime \prime \prime
}\right) -\frac{2q^{2}}{3c^{5}}\left( \eta _{R}^{i\prime \prime }\eta
_{R}^{i\prime \prime }+\eta _{I}^{i\prime \prime }\eta _{I}^{i\prime \prime
}\right),
\end{equation}%
\begin{equation}\label{PL}
P^{i\prime }=\frac{2G}{15c^{6}}\left( Q_{\mathrm{Spin}}^{kj\prime \prime
\prime }Q_{\mathrm{Mass}}^{lj\prime \prime \prime }-Q_{\mathrm{Mass}%
}^{kj\prime \prime \prime }Q_{\mathrm{Spin}}^{lj\prime \prime \prime
}\right) \epsilon _{lki}+\frac{q^{2}}{3c^{4}}\left( \eta _{I}^{k\prime
\prime }\eta _{R}^{j\prime \prime }-\eta _{R}^{k\prime \prime }\eta
_{I}^{j\prime \prime }\right) \epsilon _{jki}.
\end{equation}

\subsection{Interpretations}

\ There are a variety of comments to be made about the physical content
contained in the above kinematic and dynamical relations:

\begin{itemize}
\item A subtle (but not essential) comment for completeness should be made. We have used the symbol $q\eta^{i}$ for the complex electromagnetic dipole and $\eta ^{a}\ $for the complex center of charge. The two $\eta $s\ are related by a non-linear
term, given explicitly by (\ref{result.II})). Rather than go into a detailed explanation of our usage which could be confusing, we note that the effect is that a small quadratic term in $J_{T}^{i}$ is missing.

\item The first term of $P^{i}$ Eq. (\ref{momEM}), is the standard Newtonian kinematic
expression for the linear momentum, $M_{B}\xi _{R}^{k\,\prime }$.  This is followed by a spin-momentum coupling term of the form $(\overrightarrow{S}\times\overrightarrow{P})^{\prime}$.

\item The further term, $-\frac{2}{3}c^{-3}q^{2}\eta _{R}^{i\,\prime \prime }$%
, which is a contribution to the linear momentum from the second derivative
of the electric dipole moment, $q\eta _{R}^{i}$, plays a special role for
the case when the complex center of mass \textit{coincides} with the complex
center of charge, $\eta ^{a}=\xi ^{a}$. In this case, the second term is
exactly the contribution to the momentum that yields the classical radiation
reaction force of classical electrodynamics, \cite{LL-62}%
\begin{equation}
\frac{2}{3}c^{-3}q\xi _{R}^{i\,\prime \prime \prime }.  \label{rad.react}
\end{equation}%
In this special case we have a rather attractive identification: since now
the magnetic dipole moment is given by $D_{M}^{i}=q\xi _{I}^{i}$ and the
spin by $S^{i}=M_{B}c\xi _{I}^{i}$, we have that the gyromagnetic ratio is 
\begin{equation*}
\gamma=\frac{|S^{i}|}{|D_{M}^{i}|}=\frac{M_{B}c}{q},
\end{equation*}%
leading to the Dirac value of the $g$-factor, i.e., $g=2$.

\item Many of the remaining terms in $P^{i}$, though apparently second
order, are really of higher order when the dynamics are considered. Others involve quadrupole interactions, which
contain high powers of $c^{-1}$.

\item The complex electromagnetic dipole moment, given in general by
\begin{equation*}
D_{E\& M}^{i}=q(\eta _{R}^{i}+i\eta _{I}^{i})=D_{E}^{i}+iD_{M}^{i},
\end{equation*}
becomes, when the world-lines coincide, $D_{E\& M}^{i}=q(\xi
_{R}^{i}+i\xi _{I}^{i}).$

\item In the expression for $J^{i}$ (\ref{J*}) we have already identified, in the
earlier discussion, the first two terms, $S^{i}=M_{B}c\xi _{I}^{j}$ and $%
M_{B}\xi _{R}^{k\,\prime }\xi _{R}^{i}\epsilon _{ikj}$ as the intrinsic spin
angular momentum and the orbital angular momentum (i.e., $\overrightarrow{r}\times \overrightarrow{P}$) respectively. An interesting
contribution to the total angular momentum comes from the term, $\frac{2}{3}
c^{-2}q^{2}\eta _{I}^{i\,\prime }=\frac{2}{3}c^{-2}qD_{M}^{i\,\prime }$,
i.e., a contribution to the total angular momentum from a time-varying
magnetic dipole. A question arises: is this an observable prediction?

\item Our identification of $J^{i}$ as the total angular momentum in the
absence of a Maxwell field agrees with most other identifications (assuming
our approximations) \cite{Szabados:2004}. Very strong support of this view, with the Maxwell terms added in, comes from the flux law. In Equation~(\ref{FluxT}) we see
that there are four flux terms (more arise if we included
electromagnetic quadrupole radiation): the first and second come from the Maxwell dipole flux, while the third and fourth are the gravitational quadrupole flux terms. The Maxwell dipole part of the flux is identical to that derived from pure Maxwell theory \cite{LL-62}.
We emphasize that this angular momentum flux law has little to do directly
with the chosen definition of angular momentum. The imaginary part
of the Bianchi identity, Equation~(\ref{AsyBI2*}), \textit{is} the
conservation law. How to identify the different terms (i.e., identifying
the time derivative of the angular momentum and the flux terms) comes from
different arguments. The identification of the Maxwell contribution to total
angular momentum and the flux contain certain arbitrary assignments: some
terms on the left-hand side of the equation, i.e., terms with a time
derivative, could have been moved onto the right-hand side and been called
``flux'' terms. Our assignments were governed by the question of what terms
appeared most naturally to be explicit time derivatives (thereby being
assigned to the time derivative of the angular momentum), or which terms
appeared to be physically more likely to be an angular momentum term.

\item The angular momentum conservation law can be considered as the
evolution equation for the imaginary part of the complex world line, i.e., $%
\xi _{I}^{i}(u_{\mathrm{ret}})$. The evolution for the real part is found
from the Bondi energy-momentum loss equation.

\item The Bondi mass, $M_{B}=-\frac{c^{2}}{2\sqrt{2}G}\Psi ^{0}$, and the
original mass of the Reissner--Nordstr\"{o}m (Schwarzschild) unperturbed
metric, $M_{\mathrm{RN}}=-\frac{c^{2}}{2\sqrt{2}G}\psi _{2}^{0\,0}$ (i.e.,
the $l=0$ harmonic of $\psi _{2}^{0}$) differ by a quadratic term in the
shear, the $l=0$ part of $\sigma^{0} \dot{\overline{\sigma }}^{0}$. This suggests
that the observed mass of an object is partially determined by its
time-dependent quadrupole moment - if it exists.

\item In the discussion of the Bondi energy loss theorem (\ref{ML}), we saw that we can
relate $\xi ^{ij}$ (i.e., the $l=2$ shear term) to the gravitational
quadrupole by 
\begin{equation}
\xi ^{ij}=(\xi _{R}^{ij}+i\xi _{I}^{ij})=\frac{\sqrt{2}G}{24c^{4}}(Q_{%
\mathrm{Mass}}^{ij\,\prime \prime }+iQ_{\mathrm{Spin}}^{ij\,\prime \prime })=%
\frac{\sqrt{2}GQ_{\mathrm{Grav}}^{ij\,\prime \prime }}{24c^{4}}.  \label{chi}
\end{equation}%
to obtain the standard quadrupole energy loss.

\item The Bondi mass loss theorem with electromagnetic dipole and quadrupole
radiation becomes 
\begin{eqnarray}
M_{B}^{\prime } &=&-\frac{G}{5c^{7}}\left( Q_{\mathrm{Mass}}^{ij\,\prime
\prime \prime }Q_{\mathrm{Mass}}^{ij\,\prime \prime \prime }+Q_{\mathrm{Spin}%
}^{ij\,\prime \prime \prime }Q_{\mathrm{Spin}}^{ij\,\prime \prime \prime
}\right)   \label{M'} \\
&&-\frac{2}{3c^{5}}\left( D_{E}^{i\,\prime \prime }D_{E}^{i\,\prime \prime
}+D_{M}^{i\,\prime \prime }D_{M}^{i\,\prime \prime }\right) -\frac{2}{%
45c^{7}}\left( Q_{E}^{ij\,\prime \prime \prime }Q_{E}^{ij\,\prime \prime
\prime }+Q_{M}^{ij\,\prime \prime \prime }Q_{M}^{ij\,\prime \prime \prime
}\right)   \notag
\end{eqnarray}%
with the first term the conventional gravitational radiation with the second
and third terms the electromagnetic dipole and quadrupole radiation loss.  The momentum loss with electromagnetic quadrupole contributions becomes:
\begin{eqnarray*}
P^{i\prime }&=&\frac{2G}{15c^{6}}\left( Q_{\mathrm{Spin}}^{kj\prime \prime
\prime }Q_{\mathrm{Mass}}^{lj\prime \prime \prime }-Q_{\mathrm{Mass}%
}^{kj\prime \prime \prime }Q_{\mathrm{Spin}}^{lj\prime \prime \prime
}\right) \epsilon _{lki}+\frac{q^{2}}{3c^{4}}\left( \eta _{I}^{k\prime
\prime }\eta _{R}^{j\prime \prime }-\eta _{R}^{k\prime \prime }\eta
_{I}^{j\prime \prime }\right) \epsilon _{jki} \\
& & +\frac{2}{135c^{6}}\left(
Q_{M}^{lj\prime \prime \prime }Q_{E}^{kj\prime \prime \prime
}-Q_{E}^{lj\prime \prime \prime }Q_{M}^{kj\prime \prime \prime }\right)
\epsilon _{lki}
\end{eqnarray*}

\item There are several things to observe and comment on concerning Eqs.~(%
\ref{newton2nd}) and~(\ref{recoil}): Returning to the case
when the complex center of mass and center of charge coincide the resulting
equations of motion for the world-line, namely:%
\begin{equation*}
M_{B}\xi _{R}^{i\,\prime \prime }+\frac{2}{3}c^{-3}q\xi _{R}^{i\,\prime
\prime \prime }.=F_{\mathrm{recoil}}^{k}-M_{B}^{\ \prime }\xi _{R}^{i\prime
}-\mathfrak{P}^{i\prime }\equiv F^{i}
\end{equation*}%
we \textit{observe and stress} that, aside from several extra terms, these
equations coincide with the Lorentz-Dirac equations of motion - this includes
the hyper-acceleration term and the mass loss term, $M_{B}^{\ \prime }\xi
_{R}^{i\prime }$. This result follows directly from the Einstein-Maxwell
equations. There \textit{was no model building} other than requiring that
the two complex world lines coincide - a strong condition, perhaps
equivalent to a point particle assumption. Furthermore, there was no mass
renormalization; the mass was simply the conventional Bondi mass as seen at
infinity. Further structures (e.g., spin) could remain. The problem of the
runaway solutions, though not solved here, is converted to the stability of
the Einstein--Maxwell equations with the \textquotedblleft
coinciding\textquotedblright\ condition on the two world lines. If the two
world lines do not coincide, i.e., the Maxwell world-line is formed by
independent data, then there is no problem of unstable behavior. This
suggests a resolution to the problem of the unstable solutions: one should
treat the source as a structured object, not a point, and centers of mass
and charge as independent quantities.  Alternatively, it might be possible
that the extra terms in the equations might stabilize the equations. It is
however hard to see how this could be demonstrated.

\item The $F_{\mathrm{recoil}}^{i}$ is the recoil force from momentum
radiation, other force terms could be considered as gravitational radiation
reaction.

\item There are alternative perturbations schemes that use variations of the
procedures used here. An example is the determination of the gravitational
radiation in a Schwarzschild or Reissner-Nordstr\"{o}m space-time induced by
a time-dependent Maxwell dipole radiation field. The physical
identifications agree with those found in the present work. For instance,
the perturbations induced by a Coulomb charge and general electromagnetic
dipole Maxwell field in a Schwarzschild background lead to energy, momentum,
and angular momentum flux relations \cite{Adamo:2008zp}: 
\begin{eqnarray}
M_{B}^{\prime } &=&-\frac{2}{3c^{5}}\left( D_{E}^{i\,\prime \prime
}D_{E}^{i\prime \prime }+D_{M}^{i\prime \prime }D_{M}^{i\,\prime \prime
}\right) ,  \label{classical results} \\
P^{i\prime } &=&\frac{1}{3c^{4}}D_{E}^{k\,\prime \prime }D_{M}^{j\,\prime
\prime }\epsilon _{kji},  \notag \\
J^{k\prime } &=&\frac{2}{3c^{3}}\left( D_{E}^{i\,\prime \prime
}D_{E}^{j\,\prime }+D_{M}^{i\,\prime \prime }D_{M}^{j\,\prime }\right)
\epsilon _{ijk},  \notag
\end{eqnarray}%
all of which agree exactly with predictions from classical field theory~\cite%
{LL-62}.
\end{itemize}

The familiarity of these results and those in the full text act as an
exhibit in favor of the physical identification methods described in this
work. That is, they act as a confirmation of the consistency of the
identification scheme.

\section{Discussion \& Conclusion}
\label{5}

We have studied and described a variety of geometric structures, all
occurring within the confines of classical special and general relativity,
which strongly resemble ideas in other areas of physics. Our
complex-conjugate method for describing a real, twisting shear-free, or
asymptotically shear-free NGC places the congruence's caustic set (in
Minkowski space, interpreted as its source) on a closed curve propagating in
real time; or in more suggestive parlance, a classical string or world-tube. The dual
description, or holomorphic method, constructs a complex NGC (a complex
light-cone congruence) whose apex is on complex world-line in either complex
Minkowski space (the shear-free case) or $\mathcal{H}$-space (the
asymptotically shear-free case). When one imposes a reality structure on the
null geodesics of this congruence (i.e., asks that they intersect the real
asymptotic boundary), the source becomes a real two-dimension (complex
1-dimensional) open world-sheet, also oddly reminiscent of string theory.

Furthermore, the role of $\mathfrak{I}_{\C}^{+}$ as the object which
interpolates between these two dual descriptions is reminiscent of the
conjectured holographic principle \cite{'tHooft:1993gx, Bousso:2002}. In particular, we can think of a real, twisting asymptotically shear-free NGC as determining data on $\mathfrak{I}_{\C}^{+}$ (i.e., the complex-conjugate construction), which in turn acts as a ``lens'' into $\mathcal{H}$-space, which serves as a virtual image space for the real
space-time, where the real twisting NGC becomes a complex, twist-free NGC.
Hence, one is tempted to refer to $\mathfrak{I}_{\C}^{+}$ as the holographic
screen for some application of the holographic principle to classical
general relativity. This should be contrasted against the most famous
application of the holographic principle: the AdS/CFT correspondence \cite{Maldacena:1997re, Witten:1998qj}. Here,
the AdS boundary acts as a holographic screen interpolating between type IIB string
theory in $AdS_{5}\times S^{5}$ and $\mathcal{N}=4$ super-Yang-Mills theory in
Minkowski space-time. It is interesting that we have found structures so
closely resembling the underlying holographic principle, but which involve
nothing like extra dimensions or supersymmetry.

Our observations also raise a series of potentially interesting questions
related to fleshing out the alluded-to connections between general
relativity and more ambitious theories. For instance, is it possible to
write down a sigma model for the embedding of the open world-sheet (the
holomorphic method) into $\mathcal{H}$-space which yields the real cuts of $%
\mathfrak{I}^{+}$? Also, can the role of the future asymptotic boundary as a
holographic screen be made any more precise? Furthermore, we have seen (with
several examples) how the geometry of the virtual image space ($\mathcal{H}$%
-space) allows us to make physical identifications from data on the
holographic screen. This indicates that our virtual image space represents
the \textquotedblleft physical information\textquotedblright\ side of some
holographic principle, in the same way as the conformal field theory (CFT)
side of the AdS/CFT correspondence.

It should be noted that in 't Hooft's original work connecting gauge theory and string theory in the planar limit, no supersymmetry was introduced \cite{'tHooft:1974}; an extra dimension for string propagation does enter for anomaly cancellation, though.  It could be possible that the analytic continuation of $\mathfrak{I}^{+}$ to the holographic screen $\mathfrak{I}^{+}_{\C}$ in our investigation serves an analogous purpose, adding the degrees of freedom necessary to solve the good cut equation and construct the virtual light cone congruence.  However, it is unclear whether further degrees of freedom (in the form of additional dimensions) would be needed if our notion of the source of ``real '' null geodesics as an open string in $\mathcal{H}$-space were taken seriously (i.e., if one attempted to quantize the theory of such ``strings'').

Other (perhaps less abstract) questions immediately arise. Since we obtain many of the standard
classical mechanics kinematic and dynamic equations, is it possible that the
standard quantization could shed light on the difficult issues of quantum
gravity? Since the material described here is closely related to Penrose's
twistor theory (via different versions of the Kerr theorem relating
shear-freeness and twistors; c.f., \cite{Newman:2006, Adamo:2009vu, Adamo:2010ey}) is there a twistorial version of the present findings? Though the material presented here deals with the apparent or ``virtual'' motion of compact sources as viewed from a great distance, can it be modified or generalized to treat interacting bodies? Can or should these
complex world-lines with dynamic evolution and spin structure be taken at
all seriously - and in what context?

\subsubsection*{\textit{Acknowledgments}}

It is our pleasure to thank Fernando Alday, Lionel Mason, and Roger Penrose for many useful comments and suggestions.  TMA acknowledges support from the NSF GRFP (USA) and Balliol College.

\appendix

\section{Tensorial Spin-$s$ Spherical Harmonics}
\label{A}

Throughout the text, we have made use of tensorial spin-$s$ spherical harmonics, $Y^{s}_{l\;i\ldots j}(\zeta,\bar{\zeta})$, to expand analytic functions on the 2-sphere.  Although these are closely related to the usual spherical harmonics familiar from other arenas of physics, the incorporation of spin may be new to some readers.  This appendix reviews the basic properties of the spin-$s$ harmonics and includes some of the Clebsch-Gordon expansions of their products used in our calculations.  For the original treatment of this material, see \cite{NSO:2006}.

Before introducing these functions though, we recall the definition of the $\eth$-operator (also used throughout the text), which acts as a spin-weighted covariant derivative on $S^2$ \cite{Eth67}.  Suppose $f:S^{2}\rightarrow\C$ has spin-weight $s$; then
\begin{equation}\label{eqn: eth}
\eth f= P^{1-s}\frac{\partial(P^{s}f)}{\partial \zeta}, \qquad \bar{\eth} f = P^{1+s}\frac{\partial (P^{-s} f)}{\partial \bar{\zeta}},
\end{equation}
where $P$ is taken to be any conformal factor on the 2-sphere (in the text, we always take $P\equiv 1+\zeta\bar{\zeta}$). We see that an application of the $\eth$-operator raises spin-weight by 1, while $\bar{\eth}$ decreases it by one. As we will see, this property allows us to move among the spin-weighted tensorial spherical harmonics simply by applying differential operators.

\subsection{Spherical Harmonics}

Some time ago, the generalization of ordinary spherical harmonics $%
Y_{lm}(\zeta ,\bar{\zeta})$ to spin-weighted functions $_{(s)}Y_{lm}(\zeta ,%
\bar{\zeta})$ (e.g., \cite{Harmonics1, Eth67}) was developed to allow for
harmonic expansions of spin-weighted functions on the sphere. We have used instead the tensorial form of these spin-weighted harmonics, the so-called tensorial spin-$s$ spherical harmonics, which are
formed by taking linear combinations of the $_{(s)}Y_{lm}(\zeta ,\bar{\zeta}%
) $ \cite{NSO:2006}:%
\begin{equation*}
Y_{l\ i...k}^{s}=\sum K_{l\text{\ }i...k\ (s)}^{s\ m}Y_{lm}, 
\end{equation*}
where the indicies obey $|s|\leq l$, and the number of spatial indicies
(i.e., $i...k$) is equal to $l$. \ Explicitly, these new spin-weighted
harmonics can be constructed from first defining $Y_{l\ i...k}^{l}$.

We impose our special 2-sphere Bondi null tetrad:
\begin{eqnarray}
\hat{l}^{a} &=&\frac{1}{\sqrt{2}P}\left( P,\ \zeta +\bar{\zeta},\ -i(\zeta -\bar{%
\zeta}),\ -1+\zeta \bar{\zeta}\right) ,  \label{Tets} \\
\hat{n}^{a} &=&\frac{1}{\sqrt{2}P}\left( P,\ -(\zeta +\bar{\zeta}),\ i(\zeta -\bar{%
\zeta}),\ 1-\zeta \bar{\zeta}\right) ,  \nonumber \\
\hat{m}^{a} &=&\frac{1}{\sqrt{2}P}\left( 0,\ 1-\bar{\zeta}^{2},\ -i(1+\bar{\zeta}%
^{2}),\ 2\bar{\zeta}\right) ,  \nonumber \\
P &\equiv &1+\zeta \bar{\zeta}.  \nonumber
\end{eqnarray}%
We then project these to their covariant duals, and take the spatial parts
to obtain the 1-form components, along with a useful additional form $%
c_{i}=l_{i}-n_{i}$:%
\begin{eqnarray}
l_{i} &=&\frac{-1}{\sqrt{2}P}\left( \zeta +\bar{\zeta},\ -i(\zeta -\bar{\zeta%
}),\ -1+\zeta \bar{\zeta}\right) ,  \label{Spacialtets} \\
n_{i} &=&\frac{1}{\sqrt{2}P}\left( \zeta +\bar{\zeta},\ -i(\zeta +\bar{\zeta}%
),\ -1+\zeta \bar{\zeta}\right) ,  \nonumber \\
m_{i} &=&\frac{-1}{\sqrt{2}P}\left( 1-\bar{\zeta}^{2},\ -i(1+\bar{\zeta}%
^{2}),\ 2\bar{\zeta}\right) ,  \nonumber \\
c_{i} &=&-\sqrt{2}i\epsilon _{ijk}m_{j}\bar{m}_{k}.  \nonumber
\end{eqnarray}%
From this, we define $Y_{l\ i...k}^{l}$ as \cite{NSO:2006}%
\begin{eqnarray}
Y_{l\ i...k}^{l} &=&m_{i}m_{j}...m_{k},  \label{Harmonics 1} \\
Y_{l\ i...k}^{-l} &=&\bar{m}_{i}\bar{m}_{j}...\bar{m}_{k}.  \nonumber
\end{eqnarray}

The other harmonics are determined by the action of the $\eth $-operator on
the harmonics we have defined in (\ref{Harmonics 1}). \ In particular, it
can be shown that%
\begin{eqnarray}
Y_{l\ i...k}^{s} &=&\bar{\eth }^{l-s}\left( Y_{l\ i...k}^{l}\right) ,
\label{Harmonics 2} \\
Y_{l\ i...k}^{-|s|} &=&\eth ^{l-s}\left( Y_{l\ i...k}^{-l}\right) . 
\nonumber
\end{eqnarray}

We now present a table of the tensorial spherical harmonics up to $l=2$,
where we truncate all of our expansions in this paper. Higher harmonics
can be found in \cite{NSO:2006}.

\begin{itemize}
\item $l=0$%
\[
Y_{0}^{0}=1 
\]

\item $l=1$%
\begin{eqnarray*}
Y_{1i}^{1} &=&m_{i} \\
Y_{1i}^{0} &=&-c_{i} \\
Y_{1i}^{-1} &=&\bar{m}_{i}
\end{eqnarray*}

\item $l=2$%
\[
\begin{array}{ccc}
Y_{2ij}^{2}=m_{i}m_{j} & Y_{2ij}^{1}=-\left( c_{i}m_{j}+m_{i}c_{j}\right) & 
Y_{2ij}^{0}=3c_{i}c_{j}-2\delta _{ij} \\ 
Y_{2ij}^{-2}=\bar{m}_{i}\bar{m}_{j} & Y_{2ij}^{-1}=-\left( c_{i}\bar{m}_{j}+%
\bar{m}_{i}c_{j}\right) & 
\end{array}%
\]
\end{itemize}

In addition, it is useful to give the explicit relations between these
different harmonics in terms of the $\eth $-operator and its conjugate. \
Indeed, we can see generally that applying $\eth $ once raises the spin
index by one, and applying $\bar{\eth }$ lowers the index by one. \ This in
turn means that 
\begin{eqnarray*}
\eth Y_{l\ i...k}^{l} &=&0, \\
\bar{\eth }Y_{l\ i...k}^{-l} &=&0.
\end{eqnarray*}%
Other relations for $l\leq 2$ are given by:%
\begin{eqnarray*}
\bar{\eth }Y_{1i}^{1} &=&Y_{1i}^{0}=\eth Y_{1i}^{-1} \\
\eth Y_{1i}^{0} &=&-2Y_{1i}^{1} \\
\bar{\eth }Y_{1i}^{0} &=&-2Y_{1i}^{-1}
\end{eqnarray*}

\begin{eqnarray*}
\bar{\eth }Y_{2ij}^{2} &=&Y_{2ij}^{1} \\
\bar{\eth }^{2}Y_{2ij}^{2} &=&Y_{2ij}^{0} \\
\eth Y_{2ij}^{0} &=&-6Y_{2ij}^{1} \\
\eth Y_{2ij}^{1} &=&-4Y_{2ij}^{2}
\end{eqnarray*}

Finally, due to the non-linearity of the theory, throughout this review we
have been forced to consider products of the tensorial spin-$s$ spherical
harmonics while expanding non-linear expressions. \ These products can be
expanded as a linear combination of individual harmonics using
Clebsch-Gordon expansions. \ The explicit expansions for products of
harmonics with $l=1$ or $l=2$ are given below (we omitt higher products due
to the complexity of the expansion expressions). Further products can be
found in \cite{NSO:2006}.

\subsection{Clebsch-Gordon Expansions}

\begin{itemize}
\item $l=1$ with $l=1$%
\begin{eqnarray*}
Y_{1i}^{1}Y_{1j}^{0} &=&\frac{i}{\sqrt{2}}\epsilon _{ijk}Y_{1k}^{1}+\frac{1}{%
2}Y_{2ij}^{1} \\
Y_{1i}^{1}Y_{1j}^{-1} &=&\frac{1}{3}\delta _{ij}-\frac{i\sqrt{2}}{4}\epsilon
_{ijk}Y_{1k}^{0}-\frac{1}{12}Y_{2ij}^{0} \\
Y_{1i}^{0}Y_{1j}^{0} &=&\frac{2}{3}\delta _{ij}+\frac{1}{3}Y_{2ij}^{0}
\end{eqnarray*}

\item $l=1$ with $l=2$%
\begin{eqnarray*}
Y_{1i}^{1}Y_{2ij}^{2} &=&Y_{3ijk}^{3} \\
Y_{1i}^{0}Y_{2jk}^{0} &=&-\frac{4}{5}\delta _{kj}Y_{1i}^{0}+\frac{6}{5}%
\left( \delta _{ij}Y_{1k}^{0}+\delta _{ik}Y_{1j}^{0}\right) +\frac{1}{5}%
Y_{3ijk}^{0} \\
Y_{1i}^{1}Y_{2jk}^{0} &=&\frac{2}{5}Y_{1i}^{1}\delta _{jk}-\frac{3}{5}%
Y_{1j}^{1}\delta _{ik}-\frac{3}{5}Y_{1k}^{1}\delta _{ij}+\frac{i}{\sqrt{2}}%
\left( \epsilon _{ikl}Y_{2jl}^{1}+\epsilon _{ijl}Y_{2kl}^{1}\right) +\frac{2%
}{5}Y_{3ijk}^{1} \\
Y_{1i}^{1}Y_{2jk}^{1} &=&-\frac{1}{6}\eth \left( Y_{1i}^{1}Y_{2jk}^{0}\right)
\\
Y_{2ij}^{-1}Y_{1k}^{1} &=&\frac{3}{10}Y_{1i}^{0}\delta _{jk}+\frac{3}{10}%
Y_{1j}^{0}\delta _{ik}-\frac{1}{5}Y_{1k}^{0}\delta _{ij}+\frac{i\sqrt{2}}{12}%
\left( \epsilon _{jkl}Y_{2il}^{0}+\epsilon _{ikl}Y_{2lj}^{0}\right) -\frac{1%
}{30}Y_{3ijk}^{0} \\
Y_{1i}^{0}Y_{2jk}^{1} &=&-\frac{2}{5}Y_{1i}^{1}\delta _{jk}+\frac{3}{5}%
Y_{1j}^{1}\delta _{ik}+\frac{3}{5}Y_{1k}^{1}\delta _{ij}-\frac{i}{3\sqrt{2}}%
\left( \epsilon _{ikl}Y_{2jl}^{1}+\epsilon _{ijl}Y_{2kl}^{1}\right) +\frac{4%
}{15}Y_{3ijk}^{1} \\
Y_{2ij}^{1}Y_{1k}^{-1} &=&\frac{3}{10}Y_{1i}^{0}\delta _{jk}+\frac{3}{10}%
Y_{1j}^{0}\delta _{ik}-\frac{1}{5}Y_{1k}^{0}\delta _{ij}-\frac{i\sqrt{2}}{12}%
\left( \epsilon _{jkl}Y_{2il}^{0}+\epsilon _{ikl}Y_{2lj}^{0}\right) -\frac{1%
}{30}Y_{3ijk}^{0} \\
Y_{2ij}^{2}Y_{1k}^{0} &=&\eth \left( Y_{2ij}^{2}Y_{1k}^{-1}\right)
\end{eqnarray*}
\ 
\end{itemize}

\bibliography{holo}
\bibliographystyle{JHEP}
\end{document}